\documentclass[sigconf,nonacm]{acmart}
\usepackage{placeins}
\usepackage{enumitem}
\usepackage{multirow}
\usepackage{graphicx}
\usepackage{subcaption}
\usepackage{balance}

\newtheorem{definition}{Definition}

\DeclareMathOperator{\Tr}{Tr}
\AtBeginDocument{%
  }

\copyrightyear{2027}
\acmYear{2027}
\setcopyright{cc}
\setcctype{by}
\acmConference[WSDM '27]{Proceedings of the Twentieth ACM International Conference on Web Search and Data Mining}{February 15--19, 2027}{Hong Kong, Hong Kong}
\acmBooktitle{Proceedings of the Twentieth ACM International Conference on Web Search and Data Mining (WSDM '27), February 15--19, 2027, Hong Kong, Hong Kong}
\acmDOI{10.1145/3862322.3863483}
\acmISBN{979-8-4007-3117-4/2027/02}

\begin{document}

\title{Rethinking Semantic Alignment in LLM-Enhanced Collaborative Filtering: A Spectral Decoupling Approach}


\author{Yedong Jin}
\orcid{0009-0004-9110-4960}
\email{jin.yedong.ke0@naist.ac.jp}
\affiliation{%
  \institution{Nara Institute of Science and Technology}
  \city{Ikoma}
  \country{Japan}
}

\author{Shaowen Peng}
\orcid{0000-0003-4020-9100}
\authornote{Corresponding author.}
\email{peng.shaowen@naist.ac.jp}
\affiliation{%
  \institution{Nara Institute of Science and Technology}
  \city{Ikoma}
  \country{Japan}
}

\author{Tsunenori Mine}
\orcid{0000-0002-7462-8074}
\email{mine@ait.kyushu-u.ac.jp}
\affiliation{%
  \institution{Kyushu University}
  \city{Fukuoka}
  \country{Japan}
}

\author{Shoko Wakamiya}
\orcid{0000-0002-9371-1340}
\email{wakamiya@is.naist.jp}
\affiliation{%
  \institution{Nara Institute of Science and Technology}
  \city{Ikoma}
  \country{Japan}
}

\author{Eiji Aramaki}
\orcid{0000-0003-0201-3609}
\email{aramaki@is.naist.jp}
\affiliation{%
  \institution{Nara Institute of Science and Technology}
  \city{Ikoma}
  \country{Japan}
}

\renewcommand{\shortauthors}{Yedong Jin, Shaowen Peng, Tsunenori Mine, Shoko Wakamiya, \& Eiji Aramaki}

\begin{abstract}
Recent advances in large language models (LLMs) have enabled recommender systems to incorporate rich textual representations. Existing approaches commonly align these semantic representations with collaborative embeddings in a shared latent space, yet how such alignment affects the information encoded by LLM embeddings remains insufficiently understood. In this work, we revisit LLM-enhanced recommendation from a spectral perspective and show that collaborative and semantic signals benefit from different parts of their respective spectra. Specifically, while collaborative representations are typically dominated by smooth low-frequency components due to user--item interaction homophily, semantic embeddings contain useful non-principal singular components. Through component-wise evaluation and training-dynamics analysis, we find that alignment increasingly concentrates learned representations in the dominant collaborative and principal semantic subspaces while reducing their overlap with non-principal semantic components. Controlled comparisons further show that non-principal components provide inconsistent gains under alignment but consistently improve performance through component-level decoupling, while full prediction-level decoupling achieves the best overall performance. These results indicate that alignment does not effectively exploit complementary non-principal semantic information. Motivated by these findings, we propose UniSpecRec (\textbf{Uni}fying \textbf{Spec}tral Signals for \textbf{Rec}ommendation), which applies signal-specific spectral filtering while preserving collaborative and semantic representations in their respective spaces. UniSpecRec combines their predictions without cross-space alignment or additional trainable parameters. Extensive experiments across multiple datasets, LLM encoders, and collaborative backbones demonstrate its effectiveness, efficiency, and generalizability. The source code and datasets are available at \url{https://github.com/ixilebone/UniSpecRec}.
\end{abstract}

\ccsdesc[500]{Information systems~Recommender systems}
\keywords{Collaborative Filtering, Large Language Models, Spectral Decoupling, Recommender Systems}

\maketitle

\section{Introduction}

Recommender systems have traditionally relied on collaborative filtering (CF), which learns user preferences from historical user-item interactions through ID-based representations~\cite{MF,MultVAE,NGCF,LightGCN}. Although effective for users and items with sufficient interactions, conventional CF methods often struggle with data sparsity and cold-start scenarios because ID-based representations provide limited semantic information. Recent studies have incorporated large language models (LLMs) into recommendation systems, leveraging their semantic knowledge and reasoning capabilities~\cite{P5,TALLRec,ChatREC,InstructRec,KAR}. These methods broadly follow two paradigms. The first uses the LLM as the recommender and injects collaborative information into the language space~\cite{P5,TALLRec,ChatREC,InstructRec,A-LLMRec}. For example, A-LLMRec~\cite{A-LLMRec} maps collaborative representations into the LLM token space through learned soft prompts. This paradigm, however, typically requires invoking or adapting the LLM during recommendation, resulting in substantial computational overhead. The second paradigm uses the LLM as a semantic enhancer while retaining a conventional recommender as the prediction backbone~\cite{RLMRec,LLMRec,KAR,AlphaRec}. LLMRec~\cite{LLMRec} augments the interaction graph with LLM-generated content, while KAR~\cite{KAR} enriches representations through knowledge-augmented reasoning. Other methods explicitly map semantic embeddings into the collaborative space: RLMRec~\cite{RLMRec} performs cross-view alignment, whereas AlphaRec~\cite{AlphaRec} projects item-text embeddings into the behavior space and optimizes them with a recommendation objective. Although efficient and effective, these methods generally treat semantic embeddings as holistic features. It therefore remains unclear which semantic components contribute to recommendation and whether they are preserved during alignment. This motivates our central question: \textbf{Is explicit semantic-to-collaborative alignment necessary for effectively exploiting LLM-derived representations?}
\par

To answer this question, we revisit LLM-enhanced recommendation from a spectral perspective and uncover two key findings. First, collaborative and semantic signals benefit from different parts of their respective spectra. Collaborative recommendation primarily relies on smooth, low-frequency components of the user-item interaction graph, whereas the utility of semantic representations extends beyond their principal singular components to non-principal components. Second, alignment-based training shifts learned representations toward the dominant collaborative and principal semantic subspaces and away from non-principal semantic directions. Controlled comparisons show that
adding non-principal components within alignment yields inconsistent
gains, whereas decoupling them consistently improves principal-only
alignment. Full prediction-level decoupling performs best overall,
suggesting that explicit cross-space alignment is unnecessary for
effective semantic integration. Together, these results indicate that
alignment does not effectively exploit useful non-principal semantic
information and motivate a fully decoupled paradigm.

Recent work has explored structured semantic transformation through whitening, selective initialization, subspace separation, and spectral reshaping~\cite{whitenrec,llminit,alphafuse,ace}. SpecTran further exploits non-principal semantic components using a learnable adapter~\cite{spectran}. These methods primarily target sequential recommendation. To examine whether their gains extend to general CF, we adapt AlphaFuse and ACE under a matched setting. Our experiments show that both underperform the base CF model, revealing a gap between semantic integration in sequential recommendation and general CF. DisCo considers general recommendation and preserves view-specific information through learned disentanglement~\cite{DisCo}. Despite these advances, existing work does not jointly examine component utility across collaborative and semantic signals, semantic component dynamics under alignment, and the necessity of learned cross-space transformation in general CF. Our work addresses these questions through component-wise and alignment-dynamics analyses. \par

Motivated by these findings, we introduce a prediction-level decoupling paradigm that preserves collaborative and semantic representations in their respective spaces. This simple strategy consistently outperforms matched counterparts that rely on parameterized semantic-to-behavior alignment. Based on this paradigm, we propose UniSpecRec (\textbf{Uni}fying \textbf{Spec}tral Signals for \textbf{Rec}ommendation), which applies signal-specific spectral filtering and combines collaborative and semantic predictions without explicit cross-space alignment. UniSpecRec introduces no additional trainable parameters and consistently outperforms matched alignment-based counterparts across multiple collaborative backbones, LLM encoders, and datasets. The contributions of this paper are summarized as follows:

\begin{itemize}[leftmargin=10pt]
\item \textbf{Spectral Analysis of Semantic and Collaborative Signals.}
We jointly analyze the graph-frequency structure of collaborative signals and the singular-component structure of LLM-derived semantic embeddings in general collaborative filtering. Our analysis shows that collaborative signals are dominated by low-frequency components, whereas useful semantic information extends beyond the principal singular subspace to non-principal components.

\item \textbf{Rethinking the Alignment-Based Learning Paradigm.}
We show that commonly used alignment objectives increasingly concentrate learned representations in dominant collaborative and semantic subspaces while reducing the retention of informative non-principal semantic components. Empirically, prediction-level decoupling consistently outperforms matched alignment-based counterparts, motivating a reconsideration of explicit cross-space alignment.

\item \textbf{UniSpecRec: A Decoupled Spectral Framework.}
Motivated by these findings, we propose UniSpecRec, a simplified framework that applies signal-specific spectral filtering while preserving semantic and collaborative representations in their respective spaces. UniSpecRec introduces no additional trainable parameters and avoids explicit semantic-to-collaborative alignment.

\item \textbf{Extensive Experimental Evaluation.}
We conduct comprehensive experiments on multiple benchmark datasets, LLM encoders, and collaborative filtering backbones, demonstrating that UniSpecRec provides consistent performance improvements, strong efficiency, and enhanced robustness over representative LLM-enhanced recommendation methods.

\end{itemize}

\section{Preliminaries}
\subsection{LLM-Enhanced Recommendation}
\label{sec:llm_rec}
Let $\mathcal{U}$ and $\mathcal{I}$ denote the sets of users and items, respectively. The interaction matrix is defined as $\mathbf{R}\in \{0, 1\}^{\left | \mathcal{U} \right | \times \left | \mathcal{I} \right |}$. The collaborative embeddings are represented as $\mathbf{E}\in\mathbb{R}^{(\left | \mathcal{U} \right | + \left | \mathcal{I}\right|)\times d_c}=\{\mathbf{E}_U, \mathbf{E}_I \}$. Given pre-existing textual features of users and items, they are transformed through frozen LLMs into semantic representations: $\mathbf{S}\in\mathbb{R}^{(\left | \mathcal{U} \right | + \left | \mathcal{I}\right|)\times d_s}=\{\mathbf{S}_U,\mathbf{S}_I\}$. The final representations are generated by incorporating semantic and collaborative signals as:
\begin{equation}
\mathbf{O}= f_{\Theta} \left( \mathbf{E},\psi\left(\mathbf{S} \right) \right),
\end{equation}
where $\psi(\cdot)$ is a projector mapping semantic representations into the behavior space: $\mathbb{R}^{d_s} \rightarrow \mathbb{R}^{d_c} $, $f(\cdot)$ is a (parameterized) function generating the final embeddings by fusing the semantic and collaborative representations. The goal is to predict unobserved interactions estimated as the inner product between the user and item representations: $\hat{r}_{ui}=\mathbf{o}_u^\top \mathbf{o}_i$. While the learning objectives of most existing methods \cite{AlphaRec,A-LLMRec,RLMRec,LLMRec} can be summarized within the above paradigm, the mechanism of how the semantic representations contribute to recommendation performance has not been fully explored. Therefore, a fundamental research question arises: \textbf{Is the existing alignment-based learning paradigm optimal?}

\subsection{Collaborative and Semantic Spectra}
\label{sec:freq_def}
User-item interactions form a bipartite graph $\mathcal{G}=(\mathcal{V},\mathcal{E})$, where $\mathcal{V}=\mathcal{U}\cup\mathcal{I}$ and $\mathcal{E}=\{(u,i)\mid r_{ui}=1\}$. Let $\mathbf{A}$ be its symmetric normalized adjacency matrix, with eigenvalue $\lambda_k\in[-1,1]$ and corresponding unit-norm eigenvector $\phi_k$.
\begin{definition}[Graph Frequency]
\label{def:graph_freq}
The variation of a signal on the graph is defined as:
\begin{equation}
{\rm TV}_{\mathbf{A}}(\phi_k)=\left\|\phi_k-\mathbf{A}\phi_k \right\|=1-\lambda_k\in[0,2].
\end{equation} 
Components with small variation are low-frequency components that vary smoothly over connected nodes, whereas those with large variation are high-frequency components that vary rapidly.
\end{definition}
Because $\mathcal{G}$ is bipartite, its graph spectrum is directly related to the singular spectrum of the normalized interaction matrix \cite{SGFCF}:
\begin{equation}
\tilde{\mathbf{R}}
=
\mathbf{U}_C
\operatorname{diag}(\boldsymbol{\sigma}_c)
\mathbf{V}_C^\top,
\qquad
\sigma_{c,1}\geq\cdots\geq\sigma_{c,r_c}\geq0.
\end{equation}
Along the positive spectral branch, principal interaction components with larger singular values correspond to smoother, lower-frequency graph modes, whereas components with smaller singular values exhibit greater graph variation. Collaborative recommendation therefore primarily benefits from principal, low-frequency components~\cite{shen2021powerful,SGFCF}. We similarly decompose the semantic representation matrix $\mathbf{S}\in\mathbb{R}^{|\mathcal{V}|\times d_s}$ as: 
\begin{equation} 
\mathbf{S} = \mathbf{U}_S \operatorname{diag}(\boldsymbol{\sigma}_s) \mathbf{V}_S^\top, \qquad \sigma_{s,1}\geq\cdots\geq\sigma_{s,r_s}>0. 
\end{equation} 
We refer to components with larger singular values as \emph{principal semantic components} and those with smaller singular values as \emph{non-principal semantic components}. Although semantic singular-component order does not define graph frequency, both decompositions describe spectral dominance: principal collaborative components capture smooth interaction structure, while principal semantic components capture the dominant variation in the semantic representations. Since many LLM-enhanced methods retain or emphasize only principal semantic components~\cite{whitenrec,llminit,alphafuse}, we ask: \textbf{Is recommendation utility concentrated in the principal components of both signals, or do non-principal semantic components provide complementary information?}

\section{Methodology}
\subsection{Component Utility Across Collaborative and Semantic Spectra}
\label{sec:spectral_discrepancy}

It has been well established that collaborative signals are dominated by low-frequency components due to the homophily effect in user-item interactions \cite{shen2021powerful,SGFCF}. In contrast, how recommendation-relevant information is distributed across the singular components of semantic representations remains unclear for CF. For this analysis, we encode items from their textual profiles and construct user profiles by prompting an LLM with users' interaction histories. Since the profiles may encode collaborative information, we repeat the analysis without user semantic embeddings, using collaborative user embeddings and item-only semantics. This control produces consistent results, as reported in
Appendix~\ref{app:item_only}. \par

To investigate this question, we partition the singular components into $B$ contiguous spectral bands according to their descending singular values, $\sigma_{s,1}\geq\sigma_{s,2}\geq\cdots\geq\sigma_{s,r}\geq0$. The first band contains the largest, principal components, while successive bands contain progressively smaller singular values, extending toward the non-principal tail where $\sigma_{s,k}\rightarrow0$:
\begin{equation}
\begin{aligned}
\mathbf{U}_S &= [\mathbf{U}_S^{(1)} \| \cdots \| \mathbf{U}_S^{(B)}], \\
\mathbf{V}_S &= [\mathbf{V}_S^{(1)} \| \cdots \| \mathbf{V}_S^{(B)}],
\end{aligned}
\end{equation}
where $\mathbf{U}_S^{(b)} \in \mathbb{R}^{(|\mathcal{U}|+|\mathcal{I}|)\times d_b}$, $\mathbf{V}_S^{(b)} \in \mathbb{R}^{d_s \times d_b}$, and $d_b = \lceil \operatorname{rank}(\mathbf{S}) / B \rceil$. We treat each component equally by removing $\text{diag}\left(\sigma_s\right)$ and define the $b$-th band component as:
\begin{equation}
\mathbf{S}^{(b)} = \mathbf{U}_S^{(b)} {\mathbf{V}_S^{(b)}}^\top,
\end{equation}
and the cumulative reconstruction up to band $B'$:
\begin{equation}
\mathbf{S}^{(1:B')} = \sum_{b=1}^{B'} \mathbf{S}^{(b)}.
\end{equation}
We begin with the principal spectral band ($B'=1$) and progressively include additional bands toward the non-principal end of the spectrum, with $B'=2,3,\ldots,B$. Following existing methods, we project each cumulatively reconstructed semantic matrix into the behavior space through a learnable linear layer $\mathbf{W}\in\mathbb{R}^{d_s\times d_c}$:
\begin{equation}
\mathbf{H}^{(1:B')} = \mathbf{S}^{(1:B')} \cdot \mathbf{W}, \quad B' \in \{1, \ldots, B\},
\end{equation}
and optimize it with BPR loss \cite{bpr}. To control for the effect of cumulative rank expansion, we apply the same procedure to the collaborative interaction matrix.

\begin{figure*}[t]
    \centering
    \begin{subfigure}{0.23\textwidth}
        \includegraphics[width=\textwidth]{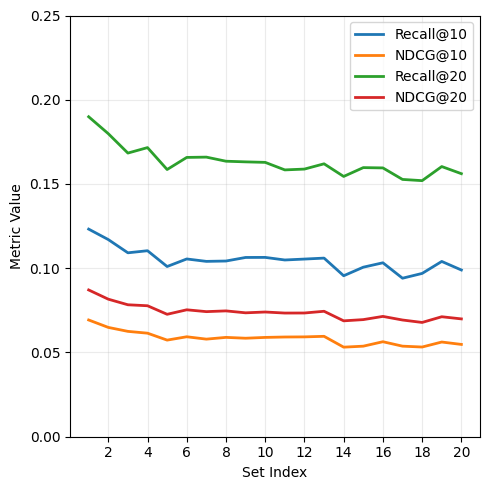}
        \caption{Interaction}
        \label{fig:sub1}
    \end{subfigure}
    \hfill
    \begin{subfigure}{0.23\textwidth}
        \includegraphics[width=\textwidth]{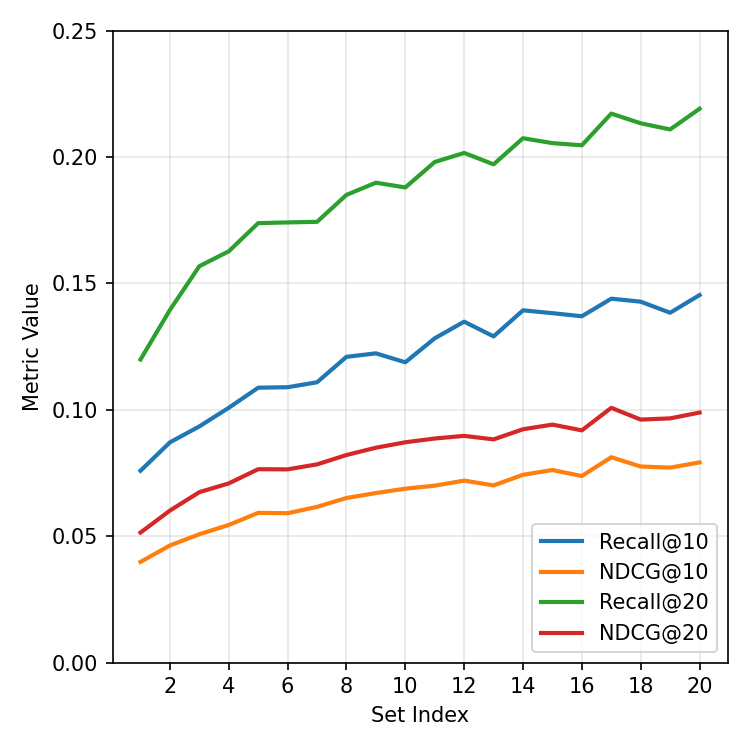}
        \caption{NV-Embed-v2}
        \label{fig:sub2}
    \end{subfigure}
    \hfill
    \begin{subfigure}{0.23\textwidth}
        \includegraphics[width=\textwidth]{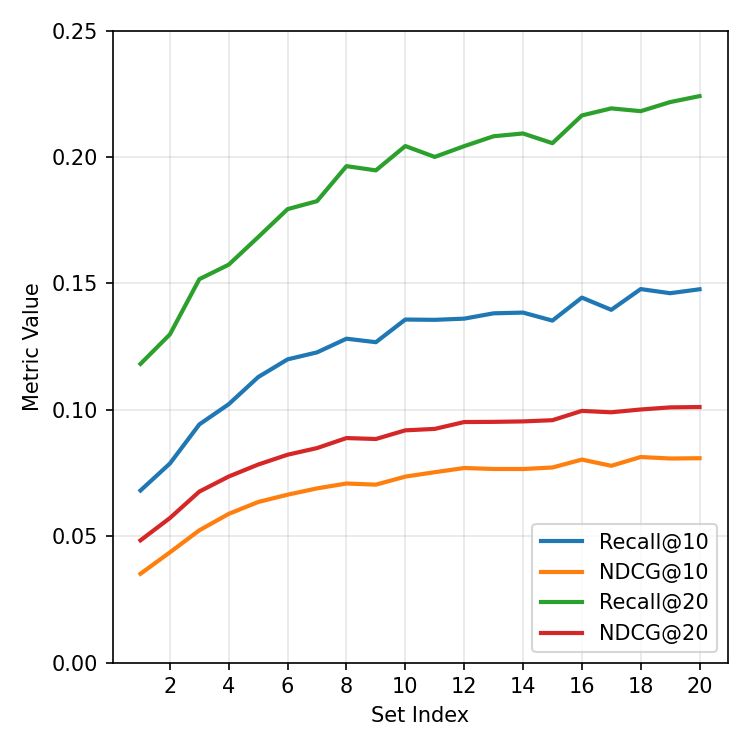}
        \caption{LLaMA-3.2-3B}
        \label{fig:sub3}
    \end{subfigure}
    \hfill
    \begin{subfigure}{0.23\textwidth}
        \includegraphics[width=\textwidth]{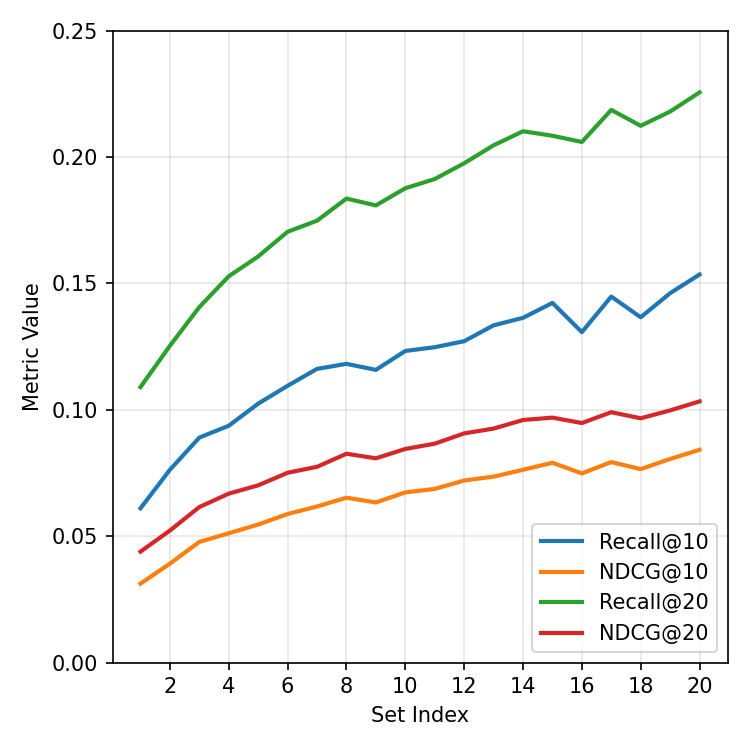}
        \caption{Qwen3-Embedding-8B}
        \label{fig:sub4}
    \end{subfigure}
    \caption{Contrasting spectral utility patterns of collaborative and semantic signals. (a) For interaction signals, performance degrades as higher graph-frequency components are included. (b)--(d) For semantic signals from three LLM encoders, performance consistently improves as bands with progressively smaller singular values are included.}
    \Description{Four plots showing recommendation performance as spectral bands are cumulatively added from principal to non-principal components. Panel (a) for the collaborative interaction matrix shows performance decreasing as higher-frequency bands are added. Panels (b), (c), and (d) for semantic embeddings from NV-Embed-v2, LLaMA-3.2-3B, and Qwen3-Embedding-8B show performance monotonically increasing as non-principal bands are included, with NV-Embed-v2 exhibiting the largest overall gain.}
    \label{fig:spectral_discrepancy}
\end{figure*}

\textbf{Observations.}
As shown in Figure~\ref{fig:spectral_discrepancy}(a), collaborative performance decreases as progressively higher graph-frequency components are included. In contrast, Figures~\ref{fig:spectral_discrepancy}(b)--(d) show that semantic performance improves monotonically as the representation expands from the principal band toward increasingly non-principal bands. Because both signals undergo the same spectral equalization and cumulative band expansion, these opposite trends cannot be explained merely by the increase in effective rank. Instead, they indicate that non-principal semantic components contain complementary information useful for recommendation. The improvement varies across encoders: NV-Embed-v2 exhibits the largest gain from the principal band to the full spectrum, followed by Qwen3-Embedding-8B and LLaMA-3.2-3B, suggesting that recommendation-relevant information is distributed differently across their singular components. Overall, collaborative recommendation primarily benefits from smooth graph structure, whereas semantic utility extends beyond the principal components.

\subsection{Semantic Component Dynamics under Alignment}
\label{sec:recap}

\begin{figure}[h]
    \centering
    \begin{subfigure}{0.23\textwidth}
        \centering
        \includegraphics[width=\textwidth]{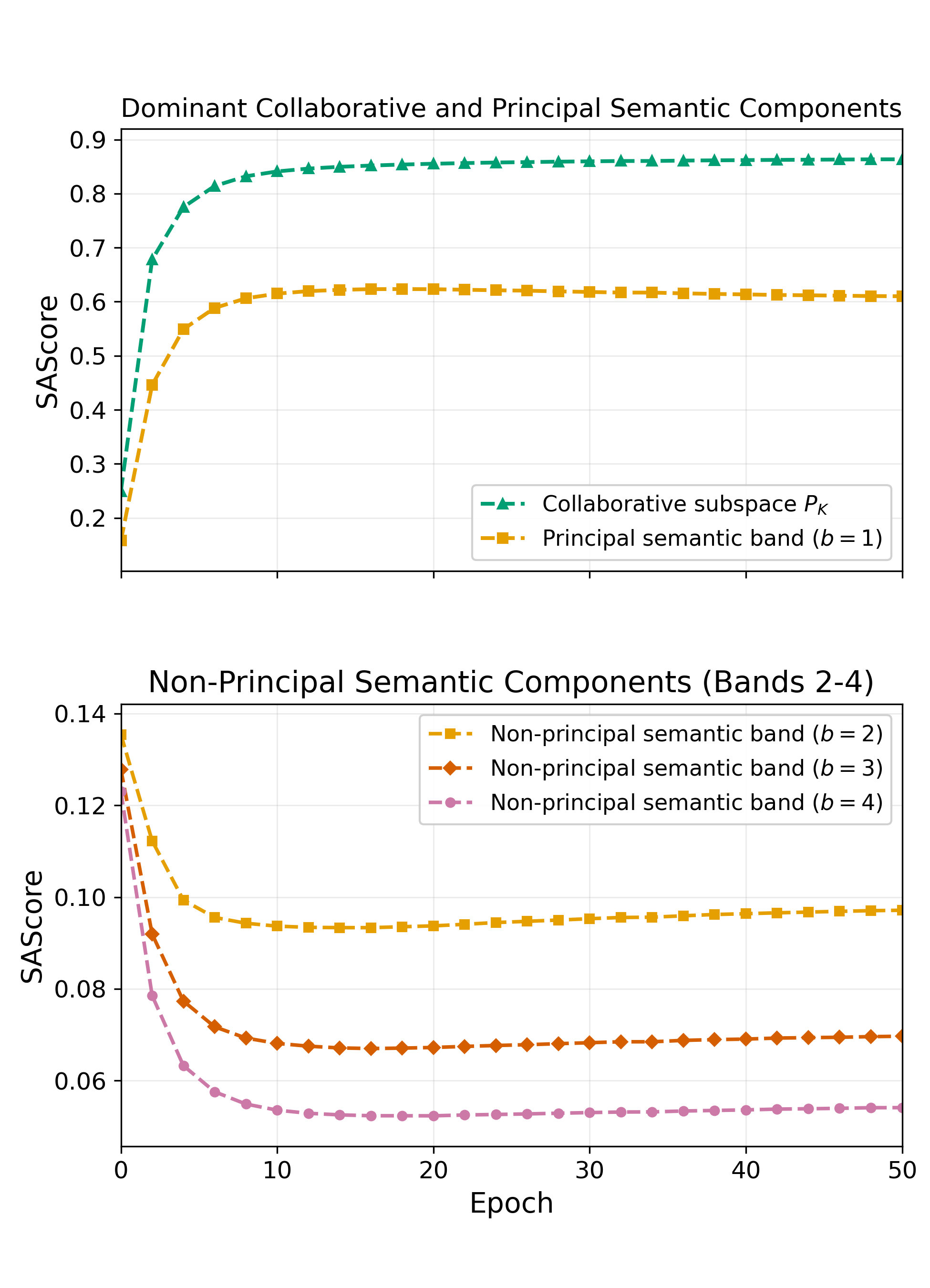}
        \subcaption{RLMRec-Gen}
        \label{subfig:rlmrecgen-epoch}
    \end{subfigure}
    \hfill
    \begin{subfigure}{0.23\textwidth}
        \centering
        \includegraphics[width=\textwidth]{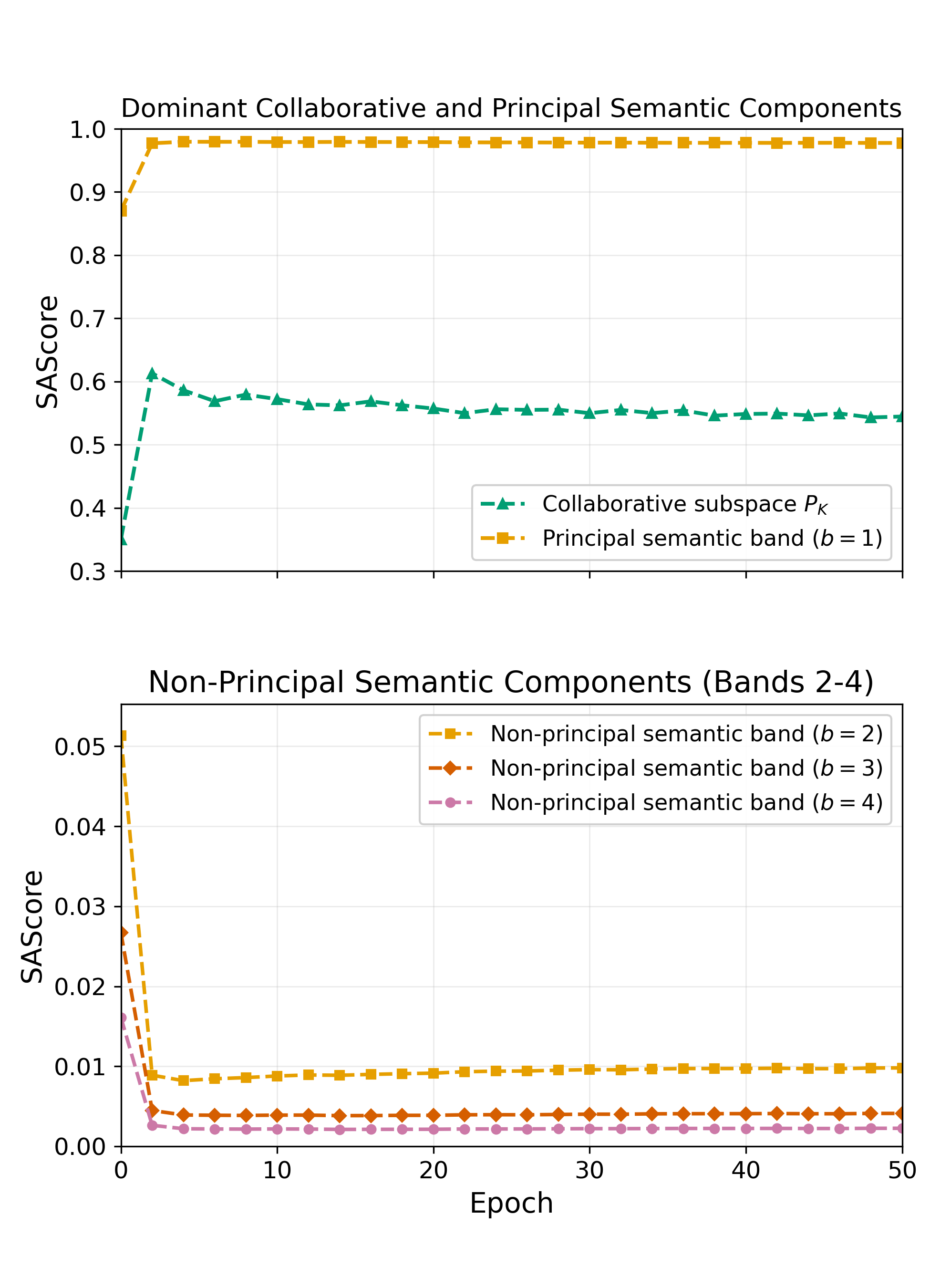}
        \subcaption{AlphaRec}
        \label{subfig:alpharec-epoch}
    \end{subfigure}
\caption{Training dynamics of SAScore under two alignment-based models.}
\Description{Two line plots of SAScore versus training epoch. The dominant collaborative and principal semantic components rises; non-principal semantic components declines.}
\label{fig:SAScore_epoch}
\end{figure}

Existing LLM-enhanced methods typically adopt alignment objectives that map semantic representations into the behavior space. A representative work AlphaRec \cite{AlphaRec} assumes a homomorphism between semantic and behavior space and replaces traditional ID embeddings with language representations initialized from $\mathbf{S}$. The optimization objective can be viewed as a low-rank approximation:
\begin{equation}
\min_{\psi} \mathcal{D} \left( \mathbf{R} , \psi(\mathbf{S}_U) \psi(\mathbf{S}_I)^\top \right) \quad 
\textit{s.t.}\operatorname{rank}(\psi(\mathbf{S}_U) \psi(\mathbf{S}_I)^\top) \le d_c
,
\end{equation}
where $\mathcal{D}(,)$ denotes a commonly used loss function measuring the difference between ground truth and predictions, and $\psi(\cdot)$ is an MLP projector mapping semantic embeddings to the behavior space. For interaction-only CF methods, the classical Eckart-Young-Mirsky \cite{EckartYoung1936} theorem shows that low-rank matrix factorization approximates the top-$d_c$ singular components of $\mathbf{R}$. In contrast, it remains unclear which components are preserved or discarded in LLM-enhanced recommendation methods. To study this, we first introduce the Spectral Alignment Score (SAScore).

\begin{definition}[Spectral Alignment Score (SAScore)]
Let $\mathbf{P} \in \mathbb{R}^{(|\mathcal{U}|+|\mathcal{I}|) \times d_P}$ be an orthonormal basis spanning a subspace satisfying $\mathbf{P}^\top\mathbf{P}=\mathbf{I}$ and $\mathbf{O} \in \mathbb{R}^{(|\mathcal{U}|+|\mathcal{I}|) \times d_c}$ denote the embeddings learned by a recommender (\textit{e.g.,} AlphaRec). The SAScore measures how much of $\mathbf{O}$ lies within the subspace spanned by $\mathbf{P}$ defined as:
\begin{equation}
\mathrm{SAScore}(\mathbf{P}, \mathbf{O}) = \frac{\left\| \mathbf{P}\mathbf{P}^\top \mathbf{O} \right\|_F^2}{\left\| \mathbf{O} \right\|_F^2}=\frac{\left\| \mathbf{P}^\top \mathbf{O} \right\|_F^2}{\left\| \mathbf{O} \right\|_F^2} \in [0, 1],
\end{equation}
where $\|\cdot\|_F$ denotes the Frobenius norm.
\end{definition}
Here, $\mathbf{P}\mathbf{P}^\top$ (\textit{i.e.,} $\mathop{\arg\min}\limits_{\mathbf{W}}\left \|\mathbf{P}\mathbf{W} - \mathbf{O} \right \|_F^2$) is the orthogonal projection onto the subspace spanned by $\mathbf{P}$, where a higher (smaller) SAScore indicates stronger (weaker) alignment. The second `=' holds as:

\begin{equation}
\left\| \mathbf{P}\mathbf{P}^\top \mathbf{O} \right\|_F^2 = \Tr\left( \left(\mathbf{P}\mathbf{P}^\top \mathbf{O}\right)^\top\left(\mathbf{P}\mathbf{P}^\top \mathbf{O}\right) \right)=\Tr\left(\left(\mathbf{P}^\top \mathbf{O}\right)^\top\left(\mathbf{P}^\top \mathbf{O}\right) \right)=\left\| \mathbf{P}^\top \mathbf{O} \right\|_F^2.
\end{equation}
In particular, we investigate how much of the learned embeddings $\mathbf{O}$ lies in (1) the optimal collaborative subspace for interaction-only CF methods spanned by:
\[
\mathbf{P}_K = \frac{1}{\sqrt{2}}
\begin{bmatrix}
\mathbf{U}_C^{(K)} \\
\mathbf{V}_C^{(K)}
\end{bmatrix}
\in \mathbb{R}^{(|\mathcal{U}|+|\mathcal{I}|)\times K},
\]
where $\mathbf{U}_C^{(K)} \in \mathbb{R}^{|\mathcal{U}| \times K}$ and $\mathbf{V}_C^{(K)} \in \mathbb{R}^{|\mathcal{I}| \times K}$ are the top-$K$ left and right singular vectors of $\mathbf{R}$, respectively (\textit{i.e.,} $\mathrm{SAScore}(\mathbf{P}_K, \mathbf{O})$); and (2) the spectral semantic subspace spanned by $\mathbf{U}_S^{(b)}$, where we set $B=4$ (\textit{i.e.,} $\mathrm{SAScore}(\mathbf{U}_S^{(b)}, \mathbf{O})$). \par
Figure~\ref{fig:SAScore_epoch} tracks SAScore throughout training for two representative alignment-based methods. In each subfigure, the upper panel reports overlap with the principal semantic band $\mathrm{SAScore}(\mathbf{U}_S^{(1)},\mathbf{O})$ and the dominant collaborative subspace $\mathrm{SAScore}(\mathbf{P}_K,\mathbf{O})$, while the lower panel reports the three non-principal semantic bands $\mathrm{SAScore}(\mathbf{U}_S^{(b)},\mathbf{O})$ for $b\in\{2,3,4\}$. For RLMRec-Gen, overlap with both the dominant collaborative subspace and principal semantic band increases rapidly and then stabilizes. For AlphaRec, overlap with the principal semantic band approaches one, whereas collaborative overlap remains moderate after an initial increase. Despite these model-specific differences, all non-principal bands decline sharply during early training and remain at low levels, with higher-index bands generally exhibiting smaller overlap. Combined with Figure~\ref{fig:spectral_discrepancy}, which establishes the recommendation utility of these components, these results indicate that alignment-based training increasingly favors principal semantic structure while failing to retain useful non-principal semantic information.

\begin{figure}[t]
    \centering
    \begin{subfigure}{0.23\textwidth}
        \includegraphics[width=\textwidth]{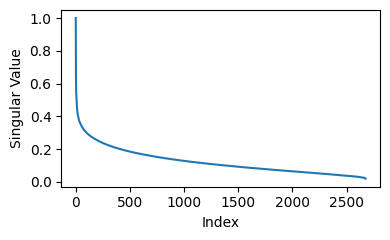}
        \caption{Interaction matrix $\mathbf{R}$}
        \Description{Line plot of singular values of the interaction matrix, showing a slowly decaying spectrum.}
        \label{fig:svd_interaction}
    \end{subfigure}
    \hfill
    \begin{subfigure}{0.23\textwidth}
        \includegraphics[width=\textwidth]{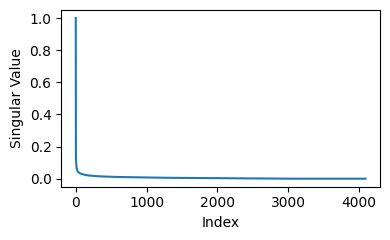}
        \caption{Semantic matrix $\mathbf{S}$}
        \Description{Line plot of singular values of the semantic matrix, showing a highly skewed spectrum with few large and many extremely small singular values.}
        \label{fig:svd_semantic}
    \end{subfigure}
\caption{Singular value distributions of the interaction matrix $\mathbf{R}$ and the semantic matrix $\mathbf{S}$.}
\label{fig:svd_distribution}
\end{figure}

Beyond the suppression induced by alignment-based training, Figure~\ref{fig:svd_distribution} reveals a strong spectral concentration in the semantic representations. Compared with the interaction matrix $\mathbf{R}$, whose normalized singular values decay relatively slowly, the semantic matrix $\mathbf{S}$ exhibits a highly skewed spectrum: a small number of principal components dominate, whereas most non-principal components have singular values several orders of magnitude smaller. Consequently, these components enter downstream models with substantially lower magnitudes and may be underutilized when $\mathbf{S}$ is directly used as input, as in many alignment-based methods. Overall, non-principal semantic information may be attenuated at two stages: first by the concentrated singular value distribution of $\mathbf{S}$ and subsequently by alignment-based training.

\subsection{Is Semantic-to-Behavior Alignment Necessary?}
\label{sec:alignment_necessary}
Motivated by these observations, we consider a \textbf{decoupling} strategy that preserves semantic and collaborative representations in their respective spaces and combines them only at the prediction level. We first apply a spectral filter $f(\cdot)$ to rebalance the contributions of non-principal components:
\begin{equation}
\begin{aligned}
\tilde{\mathbf{S}} &= \mathbf{U}_S \operatorname{diag}(f(\sigma_{s})) \mathbf{V}_S^\top, \\
\hat{\mathbf{R}}_S &= \tilde{\mathbf{S}}_U\tilde{\mathbf{S}}_I^\top =
\mathbf{U}_{S_{\left[:|\mathcal{U}|\right]}}\text{diag}(f^2(\sigma_{s}))\mathbf{U}_{S_{\left[|\mathcal{U}|:|\mathcal{U}|+|\mathcal{I}|\right]}}^\top,
\end{aligned}
\label{eq:semantic_decouple}
\end{equation}
where $\left[:|\mathcal{U}|\right]$ and $\left[|\mathcal{U}|:|\mathcal{U}|+|\mathcal{I}|\right]$ denote the corresponding user and item row partitions of $\mathbf{U}_S$, which we denote as $\mathbf{U}_{S_U}$ and $\mathbf{U}_{S_I}$ for simplicity, respectively. We use a simple spectral filter defined as $f(\sigma)=\sigma^p$, where $p \in [0, 1]$ controls the relative
contribution of different semantic components. Specifically, a smaller $p$ flattens the singular-value distribution and therefore
increases the relative contribution of non-principal components, whereas a larger $p$ places greater emphasis on principal components. Then we combine the spectrally filtered semantic prediction with collaborative signals through a simple weighted combination:
\begin{equation}
\hat{\mathbf{R}} = [\mathbf{O}_{C_U}, \mathbf{U}_{S_U}]
\begin{bmatrix}
(1-\alpha)\mathbf{I}_{d_c} &  \\
 & \alpha\operatorname{diag}(f^2(\sigma_{s}))
\end{bmatrix}
[\mathbf{O}_{C_I}, \mathbf{U}_{S_I}]^\top,
\label{eq:prototype}
\end{equation}
where $\mathbf{O}_{C_U}$ and $\mathbf{O}_{C_I}$ are user and item embeddings from a CF model, respectively; $\alpha$ is a balancing coefficient. This decoupling strategy requires no parameterized alignment module.

To examine the necessity of alignment, we compare the following baselines and controlled variants:
\begin{itemize}[leftmargin=10pt] 
\item \textbf{LightGCN}: An interaction-only CF method.

\item \textbf{AlphaRec-Full}: AlphaRec using all semantic components.

\item \textbf{AlphaRec-PC}: AlphaRec using only principal components.

\item \textbf{AlphaRec-PC+NPC-Bypass}: AlphaRec-PC with non-principal components added at the prediction level via Equation~(\ref{eq:semantic_decouple}), using only the NPCs in the semantic branch.

\item \textbf{Full Decoupling}: LightGCN and semantic predictions are computed separately and combined via Equation~(\ref{eq:prototype}).
\end{itemize}
For fair comparison, trainable variants use the same optimization objective. Table~\ref{tab:alignment_comparison} shows that semantic-enhanced variants outperform interaction-only LightGCN. AlphaRec-Full and AlphaRec-PC exhibit no consistent ordering across datasets, suggesting that adding non-principal components within the alignment pathway provides limited and inconsistent benefits. In contrast, AlphaRec-PC+NPC-Bypass consistently outperforms AlphaRec-PC and generally improves over AlphaRec-Full, demonstrating that non-principal components are useful but more effectively utilized outside alignment. Full Decoupling achieves the strongest overall performance, although its advantage on Games is modest. One possible explanation is the relatively high interaction density of Games, which provides stronger collaborative signals and leaves less room for semantic enhancement.

\begin{table}[t]
\centering
\caption{Comparison of alignment-based and decoupled approaches on Qwen3-Embedding-8B.}
\label{tab:alignment_comparison}
\resizebox{\columnwidth}{!}{
\begin{tabular}{l|cc|cc|cc}
\toprule
\multirow{2}{*}{Model} & \multicolumn{2}{c|}{Games} & \multicolumn{2}{c|}{Toys} & \multicolumn{2}{c}{Books} \\
& R@20 & N@20 & R@20 & N@20 & R@20 & N@20 \\
\midrule
LightGCN      & 0.2169 & 0.0973 & 0.0789 & 0.0357 & 0.1410 & 0.0656 \\
AlphaRec-Full  & 0.2443 & 0.1140 & 0.1018 & 0.0463 & 0.1713 & 0.0869 \\
AlphaRec-PC    & 0.2433 & 0.1109 & 0.0961 & 0.0433 & 0.1722 & 0.0867 \\
AlphaRec-PC+NPC-Bypass & 0.2445 & 0.1120 & 0.1048 & 0.0491 & 0.1795 & 0.0893 \\
Full Decoupling & \textbf{0.2472} & \textbf{0.1142} & \textbf{0.1266} & \textbf{0.0601} & \textbf{0.1869} & \textbf{0.0947} \\
\bottomrule
\end{tabular}
}
\end{table}

\subsection{UniSpecRec: Unifying Spectral Signals for Recommendation}
\label{sec:unispecrec}

We now formalize these insights into UniSpecRec (\textbf{Uni}fying \textbf{Spec}tral Signals for \textbf{Rec}ommendation), a parameter-free framework that unifies collaborative and semantic signals through a joint spectral formulation.

\textbf{Unified Spectral Formulation.}
Following the previous notations, we propose a unified prediction that integrates both signals through a joint spectral representation:
\begin{equation}
\hat{\mathbf{R}} = [\mathbf{U}_{C},\mathbf{U}_{S_{U}}]\operatorname{diag}\begin{pmatrix}(1-\alpha)f(\sigma_{c}) & \\ & \alpha f(\sigma_{s})\end{pmatrix}[\mathbf{V}_{C},\mathbf{U}_{S_{I}}]^{\top},
\end{equation}
where $f(\cdot)$ is redefined as a general spectral filter here. Due to the block-diagonal structure, this formulation naturally decomposes into two independent spectral components:
\begin{equation}
\hat{\mathbf{R}} = (1-\alpha)\underbrace{\mathbf{U}_{C}\text{diag}(f(\sigma_{c}))\mathbf{V}_{C}^\top}_{\hat{\mathbf{R}}_{\mathrm{CF}}} + \alpha\underbrace{\mathbf{U}_{S_{U}}\text{diag}(f(\sigma_{s}))\mathbf{U}_{S_{I}}^\top}_{\hat{\mathbf{R}}_S}.
\end{equation}
Collaborative and semantic signals have complementary spectral structures and therefore use signal-specific filters: $f_c(\sigma_c)$ preserves smooth collaborative components (\textit{e.g.,} through linear decay or top-$K$ truncation), while $f_s(\sigma_s)$ increases the contribution of useful non-principal semantic components. Their predictions remain separate until fusion, avoiding cross-space transformation.

\textbf{Model-Agnostic Instantiation.}
The collaborative component $\hat{\mathbf{R}}_{\mathrm{CF}}$ can be instantiated with any CF models: when paired with spectral CF methods such as SGFCF~\cite{SGFCF}, the collaborative prediction is directly obtained through spectral filtering of $\sigma_{c}$; when paired with learnable models such as LightGCN~\cite{LightGCN} or MF~\cite{MF}, $\hat{\mathbf{R}}_{\mathrm{CF}}$ is produced from the trained model. This flexibility makes UniSpecRec applicable to a wide range of recommendation backbones.

\textbf{Complexity Analysis.}
The computational overhead of UniSpecRec comes primarily from the SVD of the semantic embeddings $\mathbf{S}$. The complexity of the reduced SVD is $O((|\mathcal{U}|+|\mathcal{I}|)d_s^2)$, which is efficient since $d_s$ (the embedding dimension, typically $d_s \leq 4096$) is independent of the number of user-item interactions. The subsequent spectral filtering and prediction via the factorized form costs $O(|\mathcal{U}|K|\mathcal{I}|)$. Importantly, UniSpecRec introduces no additional learnable parameters beyond the base CF model, and the SVD computation is a one-time preprocessing step that can be performed offline. Therefore, the overall complexity remains comparable to standard CF methods, making UniSpecRec highly efficient in practice. We report the actual running time in Table~\ref{tab:efficiency}.

\section{Experiments}
\label{sec:experiments}
We organize the experiments around four questions: (1) how UniSpecRec compares with representative interaction-only and LLM-enhanced recommendation models; (2) whether the proposed UniSpecRec is inherently efficient and maintains stable performance across different semantic encoders; (3) whether decoupling semantic and collaborative signals consistently outperforms alignment across different collaborative backbones; and (4) how the spectral filter and the balancing coefficient $\alpha$ affect performance.

\subsection{Experimental Setup}

\subsubsection{Datasets and Evaluation Protocols}
We adopt the datasets used in L$^3$AE \cite{L3AE},  with detailed statistics shown in Table~\ref{tab:dataset_statistics}. The datasets are derived from the Amazon review data in three categories (Games, Toys, and Books), where only interactions with ratings larger than $3$ are retained and 10-core filtering is applied. Each dataset is split into training, validation, and test sets with a ratio of $8{:}1{:}1$. We adopt the all-item ranking protocol, where each model ranks all unobserved items for a user. Following standard practice, we report Recall@$k$ and NDCG@$k$ with $k \in \{10,20\}$.

\begin{table}[t]
  \centering
  \caption{Statistics of the datasets used in our experiments.}
  \label{tab:dataset_statistics}
  \begin{tabular}{lcccc}
    \toprule
    Dataset & \#Users & \#Items & \#Interactions & Density \\
    \midrule
    Games & 5,222 & 2,676 & 85,690 & $6.2 \times 10^{-3}$ \\
    Toys  & 14,750 & 13,358 & 250,509 & $1.3 \times 10^{-3}$ \\
    Books & 25,300 & 30,966 & 640,901 & $8.2 \times 10^{-4}$ \\
    \bottomrule
  \end{tabular}
\end{table}

\subsubsection{Baselines}
We compare UniSpecRec with two groups of methods. \textbf{Interaction-only recommendation models} include LightGCN \cite{LightGCN}, SimGCL \cite{SimGCL}, SGFCF \cite{SGFCF}. \textbf{LLM-enhanced models} include RLMRec-Con, RLMRec-Gen \cite{RLMRec}, AlphaRec \cite{AlphaRec}, Full Decoupling, ACE \cite{ace}, AlphaFuse \cite{alphafuse}, and L$^3$AE \cite{L3AE}. Full Decoupling follows the prediction-level decoupling defined in Section~\ref{sec:alignment_necessary}. ACE and AlphaFuse are originally designed for sequential recommendation; we adapt their embedding transformation schemes to the general collaborative filtering setting under the same data splits and all-item ranking protocol. In our implementation, ACE and AlphaFuse serve as item-side semantic transformation modules within the AlphaRec backbone. They use the same AlphaRec-style graph backbone, user-side construction, and training objective; only the item-side semantic transformation is replaced. SGFCF and L$^3$AE are training-free models. RLMRec-Con, RLMRec-Gen, and AlphaRec are alignment-based LLM-enhanced models. ACE and AlphaFuse apply spectral transformations to the semantic embeddings.

We adapt ACE and AlphaFuse onto the same CF backbones for a fair comparison, while omitting SpecTran since its next-item prediction objective is incompatible with static full-ranking CF.

\subsubsection{Implementation Details}
We conduct all experiments with NVIDIA A100-PCIe-40GB and AMD EPYC 7702P. For semantic encoders, we use LLaMA-3.2-3B \cite{llama3}, NV-Embed-v2 \cite{nvembed}, and Qwen3-Embedding-8B \cite{qwen3embed}. Table~\ref{tab:overall_performance} reports results with LLaMA-3.2-3B; full results are in Appendix~\ref{app:overall_performance}. Item profiles concatenate title, category, brand, and description, while user profiles aggregate up to the 50 most recent training items; embeddings are obtained via final-layer mean pooling with $\ell_2$ normalization and shared across all models. For trainable baselines, we use AdamW with learning rate $10^{-3}$, batch size $4096$ and a hidden dimension of $32$, and early stopping on validation $\text{NDCG@20}$ with a patience of 100 epochs. For UniSpecRec, we use SGFCF as the collaborative backbone, the spectral filter is instantiated as $f(\sigma)=\sigma^p$, where $p \in [0,1]$ controls the relative emphasis placed on non-principal semantic components. We first obtain the collaborative prediction matrix $\hat{\mathbf{R}}_{\mathrm{CF}}$ using SGFCF tuned to its best validation performance, then search the filter parameter $p$ and the balancing coefficient $\alpha$ over $[0,1]$.

\subsection{Comparison (RQ1-RQ2)}

\begin{table*}[tp]
  \centering
  \caption{Overall performance comparison across three datasets with LLaMA-3.2-3B. 
  ``$*$'' indicates statistical significance at $p<0.05$ for a two-tailed $t$-test against the best-performing baseline.}
  \label{tab:overall_performance}

    \centering
    \resizebox{\textwidth}{!}{%
      \begin{tabular}{l|cccc|cccc|cccc}
        \toprule
        \multirow{2}{*}{\textbf{Model}} 
        & \multicolumn{4}{c|}{\textbf{Games}} 
        & \multicolumn{4}{c|}{\textbf{Toys}} 
        & \multicolumn{4}{c}{\textbf{Books}} \\
        & R@10 & R@20 & N@10 & N@20 
        & R@10 & R@20 & N@10 & N@20 
        & R@10 & R@20 & N@10 & N@20 \\
        \midrule
        \multicolumn{13}{c}{Interaction-only recommendation models} \\
        \midrule
        LightGCN & 0.1404 & 0.2169 & 0.0770 & 0.0973 & 0.0505 & 0.0789 & 0.0282 & 0.0357 & 0.0911 & 0.1410 & 0.0523 & 0.0656 \\
        SimGCL & 0.1536 & 0.2223 & 0.0831 & 0.1016 & 0.0597 & 0.0915 & 0.0332 & 0.0417 & 0.1104 & 0.1611 & 0.0646 & 0.0788 \\
        SGFCF & 0.1873 & 0.2651 & 0.1085 & 0.1291 & 0.1005 & 0.1355 & 0.0607 & 0.0700 & 0.1743 & 0.2316 & 0.1099 & 0.1260 \\
        \midrule
        \multicolumn{13}{c}{LLM-enhanced models} \\
        \midrule
        L$^3$AE & 0.1878 & 0.2641 & 0.1083 & 0.1288 & 0.1139 & 0.1540 & 0.0674 & 0.0781 & 0.1797 & 0.2376 & 0.1137 & 0.1300 \\
        RLMRec-Con & 0.1667 & 0.2403 & 0.0911 & 0.1107 & 0.0686 & 0.1013 & 0.0388 & 0.0475 & 0.1165 & 0.1633 & 0.0692 & 0.0824 \\
        RLMRec-Gen & 0.1627 & 0.2379 & 0.0915 & 0.1114 & 0.0729 & 0.1093 & 0.0405 & 0.0502 & 0.1155 & 0.1631 & 0.0681 & 0.0814 \\
        AlphaRec & 0.1401 & 0.2108 & 0.0791 & 0.0952 & 0.0720 & 0.1050 & 0.0400 & 0.0488 & 0.1085 & 0.1578 & 0.0627 & 0.0765 \\
        ACE & 0.1589 & 0.2362 & 0.0872 & 0.1079 & 0.0627 & 0.0953 & 0.0334 & 0.0420 & 0.1067 & 0.1513 & 0.0624 & 0.0749 \\
        AlphaFuse & 0.1646 & 0.2423 & 0.0916 & 0.1126 & 0.0609 & 0.0929 & 0.0332 & 0.0417 & 0.1231 & 0.1747 & 0.0728 & 0.0872 \\
        \hline
        Full Decoupling
        & 0.1706 & 0.2443 & 0.0948 & 0.1146
        & 0.1018 & 0.1416 & 0.0595 & 0.0701
        & 0.1388 & 0.1905 & 0.0825 & 0.0970 \\
        UniSpecRec
        & \textbf{0.1922}$^{*}$
        & \textbf{0.2787}$^{*}$
        & \textbf{0.1137}$^{*}$
        & \textbf{0.1370}$^{*}$
        & \textbf{0.1155}$^{*}$
        & \textbf{0.1563}$^{*}$
        & \textbf{0.0694}$^{*}$
        & \textbf{0.0803}$^{*}$
        & \textbf{0.1816}$^{*}$
        & \textbf{0.2419}$^{*}$
        & \textbf{0.1143}
        & \textbf{0.1312}$^{*}$ \\
        \bottomrule
      \end{tabular}%
    }
\end{table*}

\begin{table}[h]
  \centering
  \caption{Efficiency comparison (seconds).}
  \label{tab:efficiency}
  \begin{tabular}{l|c|c|c}
    \toprule
    Model & Games & Toys & Books \\
    \midrule
    LightGCN & 48 & 150 & 650 \\
    RLMRec-Con & 30 & 100 & 400 \\
    RLMRec-Gen & 52 & 220 & 980 \\
    AlphaRec & 22 & 58 & 388 \\
    L$^3$AE & 2 & 17 & 162 \\
    UniSpecRec & 8 & 14 & 52 \\
    \bottomrule
  \end{tabular}
\end{table}

\subsubsection{Overall Performance}
\label{sec:overall_performance}

Table~\ref{tab:overall_performance} reports the performance across three datasets using LLaMA-3.2-3B (see Appendix~\ref{app:overall_performance} for results across all three LLM encoders). We observe the following:
\begin{itemize}[leftmargin=10pt]
\item \textbf{Interaction-Only Baselines.}
UniSpecRec consistently outperforms all interaction-only baselines. Using the best-performing baseline, SGFCF, as a reference, UniSpecRec achieves relative Recall@20 improvements of 5.1\%, 15.3\%, and 4.4\% on Games, Toys, and Books, respectively, with the LLaMA-3.2-3B encoder. These gains indicate that prediction-level decoupling captures complementary non-principal semantic information beyond the interaction signal.

\item \textbf{Alignment vs. Decoupling.}
Fixing the backbone to LightGCN, the LLM-enhanced variants differ in whether the ID embeddings are retained and in where the semantic signal enters. AlphaRec replaces the ID embeddings by a projection of the semantics and drops to 0.2108 Recall@20 on Games, below the interaction-only LightGCN (0.2169). RLMRec-Con and RLMRec-Gen retain the ID embeddings and align only through a regularizer, reaching 0.2403 and 0.2379. Full Decoupling retains them as well but fuses the two predictions at the output level, reaching 0.2443 on Games (+1.7\% over the best alignment variant) and improving by 29.6\% and 16.7\% on Toys and Books. Because the backbone and the semantic inputs are unchanged, these differences come from how the semantic signal is integrated rather than from the collaborative model.

\item \textbf{Item-Side Spectral Transformations.}
ACE reshapes the spectrum of item semantic embeddings, whereas AlphaFuse separates a selected semantic subspace and learns ID embeddings in its orthogonal complement; both are still integrated through the same AlphaRec backbone. On Games they improve over AlphaRec (0.2362 and 0.2423 vs.\ 0.2108), confirming that reshaping the semantic embeddings is beneficial, yet on Toys and Books they remain clearly below prediction-level decoupling (0.0929--0.0953 vs.\ 0.1416; 0.1513--0.1747 vs.\ 0.1905). These results indicate that item-side spectral transformation helps, but does not substitute for decoupling as the integration mechanism.

\item \textbf{Training-Free Baselines.}
L$^3$AE treats the semantic signal without any spectral processing, while UniSpecRec applies independent spectral filtering to each signal. The consistent improvement of UniSpecRec over L$^3$AE indicates that semantic embeddings benefit from tailored spectral treatment: their most discriminative information resides in non-principal directions that raw cosine similarity fails to isolate. UniSpecRec fuses the two predictions only at the output level and therefore preserves the low-frequency structure of interaction signals while retaining discriminative detail in the non-principal components of the semantic embeddings.
\end{itemize}

\begin{table}
  \centering
  \caption{Stability comparison: variance ($\times 10^{-6}$) of Recall@20 and NDCG@20 across three LLM encoders.
  Lower values indicate better robustness.}
  \label{tab:stability}

  \resizebox{\columnwidth}{!}{%
    \begin{tabular}{lcccccc}
      \toprule
      \multirow{2}{*}{Model}
      & \multicolumn{2}{c}{Games}
      & \multicolumn{2}{c}{Toys}
      & \multicolumn{2}{c}{Books} \\
      \cmidrule(lr){2-3}
      \cmidrule(lr){4-5}
      \cmidrule(lr){6-7}
      & R@20 & N@20 & R@20 & N@20 & R@20 & N@20 \\
      \midrule
      RLMRec-Gen & 40.14 & 18.02 & 29.73 & 6.54 & 20.60 & 5.90 \\
      AlphaRec & 337.84 & 114.38 & 67.45 & 12.40 & 36.50 & 21.41 \\
      ACE & 21.61 & 15.36 & 24.21 & 7.15 & 13.41 & 5.12 \\
      AlphaFuse & 35.08 & 9.39 & 39.75 & 8.94 & 35.89 & 6.27 \\
      L$^3$AE & 24.46 & 10.28 & 31.46 & 11.23 & 4.78 & 0.54 \\
      UniSpecRec & \textbf{11.26} & \textbf{5.43} & \textbf{18.84} & \textbf{6.32} & \textbf{0.10} & \textbf{0.09} \\
      \bottomrule
    \end{tabular}%
  }
\end{table}

\subsubsection{Efficiency}

Table~\ref{tab:efficiency} reports the average running time on the same hardware. For trainable models (LightGCN, RLMRec-Con, RLMRec-Gen and AlphaRec), we report the total training time until early stopping on the validation set. For training-free models (L$^3$AE and UniSpecRec), we report the total processing time. Among trainable models, AlphaRec and RLMRec-Con are the fastest and RLMRec-Gen the slowest; both training-free methods finish within a minute on Toys. UniSpecRec is $2.8\times$/$4.1\times$/$7.5\times$ faster than AlphaRec and more than $3\times$ faster than L$^3$AE on Books.

\subsubsection{Stability}
\label{sec:stability}
Learned alignment may be sensitive to the spectral geometry of the source embeddings. We quantify this sensitivity using the variance of Recall@20 and NDCG@20 across three encoders, where lower variance indicates greater stability. As shown in Table~\ref{tab:stability}, UniSpecRec achieves the lowest variance across all datasets and metrics. AlphaRec is substantially less stable, while ACE, AlphaFuse, and L$^3$AE reduce this gap but remain more sensitive than UniSpecRec. These results suggest that UniSpecRec's robustness arises from its decoupled spectral combination rather than from training-free inference alone.

\begin{table*}[t]
  \centering
  \caption{Comparison between alignment and decoupling paradigms under matched collaborative backbones on Games.}
  \label{tab:alignment_vs_decoupling}
    \begin{tabular}{l|lcc|cc|cc}
      \toprule
      \multirow{2}{*}{\textbf{Paradigm}} & \multirow{2}{*}{\textbf{Model}} & \multicolumn{2}{c|}{\textbf{Llama}} & \multicolumn{2}{c|}{\textbf{NVIDIA}} & \multicolumn{2}{c}{\textbf{Qwen}} \\
      & & R@20 & N@20 & R@20 & N@20 & R@20 & N@20 \\
      \midrule
      \multirow{3}{*}{\textbf{Alignment}} & MF      & 0.0685 & 0.0294 & 0.1130 & 0.0517  & 0.1027 & 0.0475 \\
      & LightGCN & 0.2309 & 0.1079 & 0.2478 & 0.1167 & 0.2434 & 0.1132 \\
      & SimGCL   & 0.1755 & 0.0787 & 0.2210 & 0.1000 & 0.1810 & 0.0811 \\
      \midrule
      \multirow{3}{*}{\textbf{Decoupling}} & MF      & 0.1762 & 0.0855 & 0.1954 & 0.0925 &0.1579 & 0.0761 \\
      & LightGCN & 0.2496 & 0.1183 & 0.2545 & 0.1221 & 0.2451 & 0.1143 \\
      & SimGCL   & 0.2452 & 0.1147 & 0.2589 & 0.1230 & 0.2352 & 0.1098 \\
      \bottomrule
    \end{tabular}%
\end{table*}

\subsection{Further Study of Alignment and Decoupling Paradigms (RQ3)}
\label{sec:further_decoupling}

Table~\ref{tab:alignment_comparison} showed that decoupling LightGCN and semantic predictions outperforms AlphaRec, the alignment-based method built on the same backbone. To verify that this advantage reflects a general property of the paradigm, we extend the comparison on Games to two additional backbones: MF and SimGCL. For each backbone, its semantic-enhanced variant is constructed under two paradigms:

\begin{itemize}[leftmargin=10pt]

\item \textbf{Alignment paradigm.}  
For MF, the user and item latent factors are replaced by $\psi(\mathbf{S}_U)$ and $\psi(\mathbf{S}_I)$, respectively, where $\psi(\cdot)$ is a learnable MLP.  
For LightGCN and SimGCL, the ID-based embedding table is replaced analogously.

\item \textbf{Decoupling paradigm.}  
Here, the collaborative backbone is trained independently in its original ID-based form. Its predictions are then linearly combined with the spectrally filtered semantic predictions, without any parameterized alignment module. The combination coefficient $\alpha$ and filter exponent $p$ are selected on the validation set.
\end{itemize}
Table~\ref{tab:alignment_vs_decoupling} shows that decoupling consistently outperforms alignment across all three backbones and encoders. The gap is largest for MF: under NV-Embed-v2, decoupling achieves Recall@20 of 0.1954 compared to 0.1130 for alignment. The same pattern holds for the graph-based backbones (LightGCN and SimGCL), which indicates that the advantage is not tied to a specific architecture but follows from keeping the two signals in their respective spaces (see Section~\ref{sec:alignment_necessary}).

\begin{table}
  \centering
  \caption{Comparison of different semantic filter designs.}
  \label{tab:filter_comparison}
  \resizebox{\columnwidth}{!}{
  \begin{tabular}{l c c c c c c}
    \toprule
    Filter & \multicolumn{2}{c}{Llama} & \multicolumn{2}{c}{NVIDIA} & \multicolumn{2}{c}{Qwen} \\
    \cmidrule(lr){2-3} \cmidrule(lr){4-5} \cmidrule(lr){6-7}
           & R@20 & N@20 & R@20 & N@20 & R@20 & N@20 \\
    \midrule
    Identity               & 0.2661 & 0.1295 & 0.2734 & 0.1323 & 0.2663 & 0.1300 \\
    Log                    & 0.2767 & 0.1359 & 0.2815 & \textbf{0.1382} & 0.2737 & \textbf{0.1335} \\
    Power                  & \textbf{0.2787} & \textbf{0.1370} & \textbf{0.2823} & 0.1381 & 0.2741 & 0.1327 \\
    Rational               & 0.2769 & 0.1361 & 0.2810 & 0.1381 & \textbf{0.2745} & \textbf{0.1335} \\
    Train                  & 0.2669 & 0.1306 & 0.2774 & 0.1350 & 0.2676 & 0.1293 \\
    \bottomrule
  \end{tabular}
  }
\end{table}

\subsection{Model Analysis (RQ4)}
\label{sec:model_analysis}
UniSpecRec has two key hyperparameters: the spectral filter $f(\sigma)$ that re-weights the singular values of the semantic matrix, and the balancing coefficient $\alpha$ that interpolates between the collaborative and semantic prediction matrices. Encoder-specific filter curves and the joint sensitivity of $p$ and $\alpha$ are provided in Appendix~\ref{app:param_analysis}.

\subsubsection{Analysis of Spectral Filters}
We first analyze the semantic spectral filter $f(\sigma)$. Our default choice is the power filter $f(\sigma)=\sigma^p$. We compare it against several alternatives: an identity filter (no transformation), a logarithmic filter, a rational filter, and a learnable filter optimized on the validation set. Specifically, \textbf{Power:} $f(\sigma) = \sigma^p$, where $p \in [0, 1]$ controls the degree of spectral flattening; a smaller $p$ increases the relative contribution of non-principal components. \textbf{Log:} $f(\sigma) = \log(1 + \gamma \cdot \sigma / \sigma_{\max})$, where $\gamma \in [0, 250]$ and $\sigma_{\max} = \max_k \sigma_k$, providing a gentler rescaling of the singular value spectrum. \textbf{Rational:} $f(\sigma) = \frac{\sigma / \sigma_{\max}}{\sigma / \sigma_{\max} + \beta}$, where $\beta \in [0, 0.25]$, which suppresses large singular values while preserving the relative ordering. \textbf{Train:} $f(\sigma) = \text{MLP}(\sigma)$, where a small multi-layer perceptron processes the original singular values and is optimized on the validation set using the BPR loss.

\begin{table}
  \centering
  \caption{Best validation-selected $p$ and $\alpha$ across datasets and LLM encoders.}
  \label{tab:param_analysis}
  \begin{tabular}{l c c c c c c}
    \toprule
    \multirow{2}{*}{Encoder} & \multicolumn{2}{c}{Games} & \multicolumn{2}{c}{Toys} & \multicolumn{2}{c}{Books} \\
    \cmidrule(lr){2-3} \cmidrule(lr){4-5} \cmidrule(lr){6-7}
    & $p$ & $\alpha$ & $p$ & $\alpha$ & $p$ & $\alpha$ \\
    \midrule
    LLaMA-3.2-3B      & 0.27  & 0.35 & 0.15 & 0.40  & 0.22 & 0.34 \\
    NV-Embed-v2       & 0.58 & 0.40  & 0.48 & 0.42  & 0.10  & 0.30  \\
    Qwen3-Embedding-8B& 0.58 & 0.21 & 0.53  & 0.38 & 0.21  & 0.24 \\
    \bottomrule
  \end{tabular}
\end{table}

\begin{figure}[t] 
    \centering 
    \includegraphics[width=\linewidth]{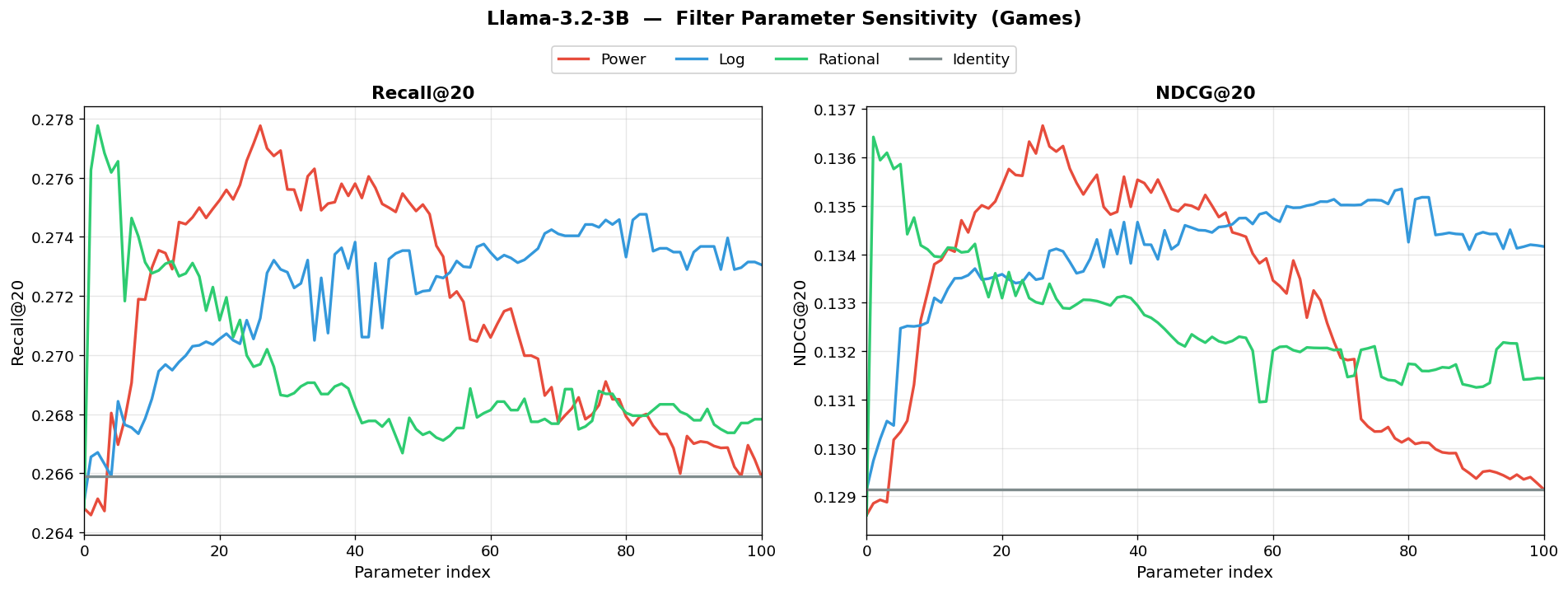}

    \caption{Performance comparison of different spectral filter designs on Games with LLaMA-3.2-3B.}
    \Description{Line plots comparing recommendation performance across hyperparameter ranges for different spectral filters on the Games dataset with LLaMA-3.2-3B. The power filter maintains a relatively flat and high performance curve across its parameter range from 0 to 1, whereas the logarithmic and rational filters exhibit substantially larger performance variations across their parameter ranges.}
    \label{fig:spectral_filters}
\end{figure}

Table~\ref{tab:filter_comparison} and Figure~\ref{fig:spectral_filters} yield two main observations. First, all non-identity hand-crafted filters outperform the learnable MLP, suggesting that explicit spectral reweighting is more effective than unconstrained transformation in this setting. The identity filter performs worst, indicating that directly preserving the raw singular-value spectrum is suboptimal. Second, the power filter is more robust to its parameter than the logarithmic and rational filters. As shown in Figure~\ref{fig:spectral_filters}, the performance of the latter two varies considerably across their parameter ranges, whereas the power filter exhibits a relatively flat performance curve over the bounded interval $p\in[0,1]$. Consequently, a competitive value of $p$ can be identified through a coarse search. As reported in Table~\ref{tab:param_analysis}, the optimal $p$ varies moderately across datasets and encoders, typically within $[0.1,0.6]$. This range supports partial compensation for the sharp decay of the semantic spectrum, strengthening informative non-principal components without excessively amplifying noise.

\subsubsection{Analysis of \texorpdfstring{$\alpha$}{alpha}}
We now analyze the balancing coefficient $\alpha$, which controls the relative contribution of the semantic branch in the decoupled predictor $\hat{\mathbf{R}} = (1 - \alpha) \hat{\mathbf{R}}_{\mathrm{CF}} + \alpha \hat{\mathbf{R}}_S$.
We analyze its behavior jointly with the power filter exponent $p$ to understand how the two hyperparameters interact. Table~\ref{tab:param_analysis} shows a clear contrast between the two hyperparameters. The optimal $p$ varies across encoders and datasets (from $0.10$ to $0.58$), but the optimal $\alpha$ stays within a narrow interval $[0.21, 0.42]$ and clusters around $0.35$ across all nine configurations. This holds despite large differences in dataset sparsity: Games has density $6.2 \times 10^{-3}$ while Books is an order of magnitude sparser at $8.2 \times 10^{-4}$. In practice, this means $\alpha$ can be set to $\approx 0.35$ without per-dataset grid search; a coarse sweep over $\{0.25, 0.30, 0.35, 0.40\}$ is enough to locate a near-optimal value.

\section{Related Work}

\textbf{Collaborative Filtering and Spectral Graph Methods.}
Collaborative Filtering (CF) is the foundation of recommender systems, with early work spanning matrix factorization~\cite{MF,SVD++}, neural approaches~\cite{NeuMF}, and pairwise ranking~\cite{bpr}, and later graph neural networks such as NGCF~\cite{NGCF} and LightGCN~\cite{LightGCN} that capture high-order connectivity through neighborhood aggregation. The performance of graph-based CF can be understood from a spectral perspective: the homophily effect concentrates collaborative signals in low-frequency components, while high-frequency components are treated as noise~\cite{shen2021powerful,SGFCF}. Recent works revisit this low-pass assumption through truncated SVD~\cite{SVDGCN}, Chebyshev interpolation for non-linear filtering~\cite{ChebyCF}, and bi-level optimization for adaptive filter learning~\cite{ASPIRE}. These advances focus exclusively on the collaborative interaction graph and do not consider external semantic signals, which is the focus of this work.

\noindent\textbf{LLM-Enhanced Recommendation.}
Large Language Models have been integrated into recommender systems to address data sparsity and cold-start challenges~\cite{LLM4RecSurvey}. Existing approaches can be broadly organized along two dimensions. The first treats the LLM as a recommender, injecting collaborative knowledge into the language space. Early efforts in this direction adopt prompting or fine-tuning strategies to align LLMs with recommendation~\cite{P5,TALLRec,InstructRec,ChatREC}. Subsequent work refines this paradigm through more sophisticated alignment mechanisms, including soft-prompt projection of collaborative embeddings~\cite{A-LLMRec}, data-efficient fine-tuning~\cite{DEALRec}, multi-facet item identifiers~\cite{TransRec}, textual ID learning for zero-shot recommendation~\cite{IDGenRec}, and curriculum-based hybrid prompting~\cite{LLaRA}. The second paradigm uses the LLM as an enhancer, mapping semantic representations into the behavior space. Representative methods augment the interaction graph with LLM-generated content~\cite{LLMRec,KAR}, or directly project semantic embeddings into the behavior space through cross-view contrastive alignment~\cite{RLMRec} or MLP projectors trained with recommendation objectives~\cite{AlphaRec}. CoLLM~\cite{CoLLM} treats collaborative information as a separate modality and integrates it through a dedicated mapping module. More recently, generative recommendation~\cite{EAGER,ETEGRec} represents items as semantic tokens and generates recommendations autoregressively, while agent-based~\cite{AgentCF} and meta-item embedding~\cite{MI4Rec} approaches explore orthogonal directions for cold-start scenarios. Despite the diversity of these methods, most alignment-based methods treat semantic embeddings as holistic features without examining how individual spectral components are affected during semantic-to-collaborative transformation.

\noindent\textbf{Spectral and Decoupled Semantic Integration.} Several methods exploit the structure of language embeddings for recommendation. WhitenRec applies whitening, LLMInit selectively initializes collaborative representations, AlphaFuse learns ID embeddings in the semantic null space, and ACE reshapes the semantic spectrum~\cite{whitenrec,llminit,alphafuse,ace}. These methods are designed primarily for sequential recommendation and transform item embeddings before sequence modeling; their gains therefore do not directly establish effectiveness in general CF. Indeed, our adapted AlphaFuse and ACE variants fail to consistently improve over the base CF model. SpecTran demonstrates the utility of non-principal components, but aggregates them into a transformed item representation through a learnable adapter rather than separating semantic and collaborative predictions~\cite{spectran}. In general CF, DisCo preserves shared and view-specific information through learned disentanglement, while L$^3$AE combines EASE-based collaborative predictions with unfiltered semantic similarity~\cite{DisCo,L3AE}. Unlike these methods, UniSpecRec applies signal-specific spectral filtering and combines predictions without a learned alignment or disentanglement module.

\section{Conclusion}
In this work, we revisited LLM-enhanced recommendation from a spectral perspective. We found that different spectral components matter for collaborative and semantic signals: collaborative recommendation mainly relies on smooth low-frequency components, whereas non-principal semantic components also contribute to recommendation accuracy. We further showed that prevailing alignment-based objectives implicitly favor low-rank representations, thereby suppressing informative non-principal semantic components. Building on this insight, we proposed UniSpecRec, a spectral decoupling framework that keeps collaborative and semantic signals in their respective domains and fuses them at the prediction level. Extensive experiments showed that UniSpecRec consistently outperforms alignment-based methods and achieves better stability across different LLM encoders and datasets. For future work, we plan to investigate what information is encoded by non-principal semantic components and how singular-component structure relates to semantic granularity. We also plan to extend the decoupling principle to broader signal-fusion settings, such as multimodal recommendation.

\section*{Acknowledgement}
This work was supported in part by JSPS KAKENHI (Grant No. 25K21353) and the Cross-Ministerial Strategic Innovation Promotion Program (SIP) on `Integrated Health Care System' (Grant No. JPJ012425).

\section*{Ethical Considerations}

This work uses publicly available and anonymized recommendation benchmarks and does not involve direct data collection from individuals. Nevertheless, recommendation systems may reinforce popularity and exposure biases, while user-history-based semantic representations may reveal sensitive preferences when applied to identifiable data. The pretrained language encoders may also inherit biases from their training data. Our evaluation is conducted offline and does not assess privacy leakage, fairness across user groups, or downstream societal effects. Any practical deployment should apply data minimization and access controls, obtain appropriate consent, and audit recommendation quality and exposure across relevant user and item groups.

\bibliographystyle{ACM-Reference-Format}
\balance
\bibliography{sample-base}
\clearpage
\nobalance
\appendix

\section{Item-only Control}
\label{app:item_only}
\noindent\textit{Complementing Section~\ref{sec:spectral_discrepancy}.}

\begin{figure*}[ht]
    \centering
    \begin{subfigure}{0.32\textwidth}
        \includegraphics[width=\textwidth]{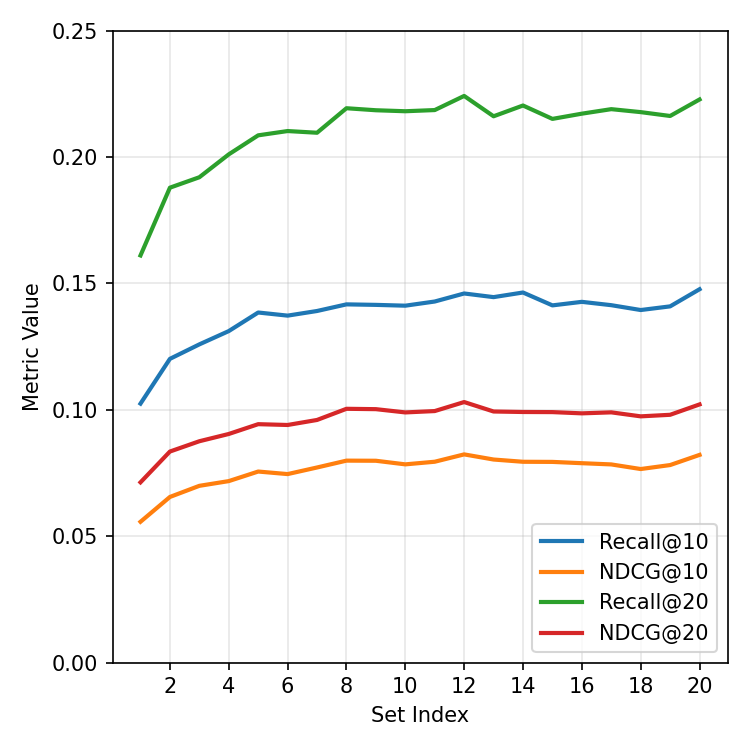}
        \caption{Games, NV-Embed-v2}
        \label{fig:app_item_sub1}
    \end{subfigure}
    \hfill
    \begin{subfigure}{0.32\textwidth}
        \includegraphics[width=\textwidth]{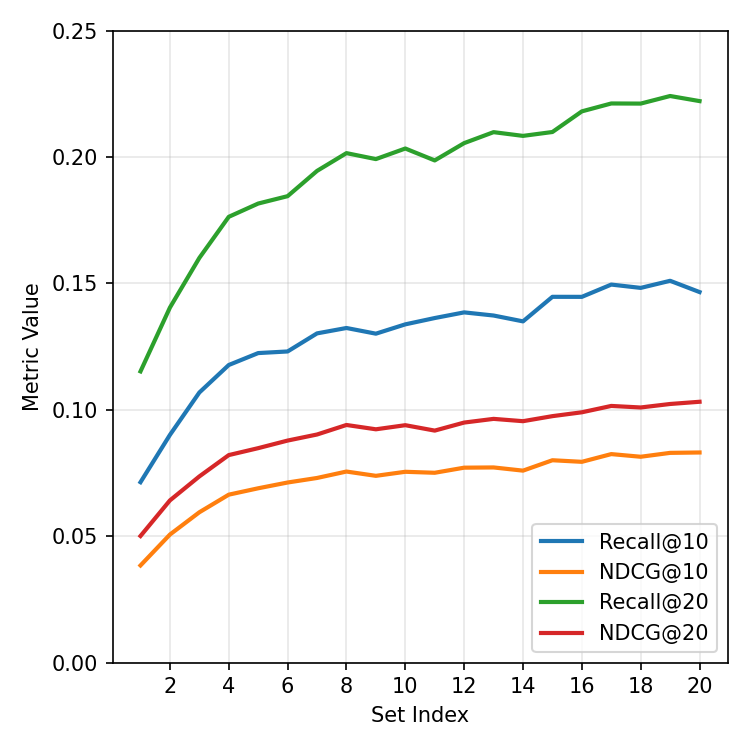}
        \caption{Games, LLaMA-3.2-3B}
        \label{fig:app_item_sub2}
    \end{subfigure}
    \hfill
    \begin{subfigure}{0.32\textwidth}
        \includegraphics[width=\textwidth]{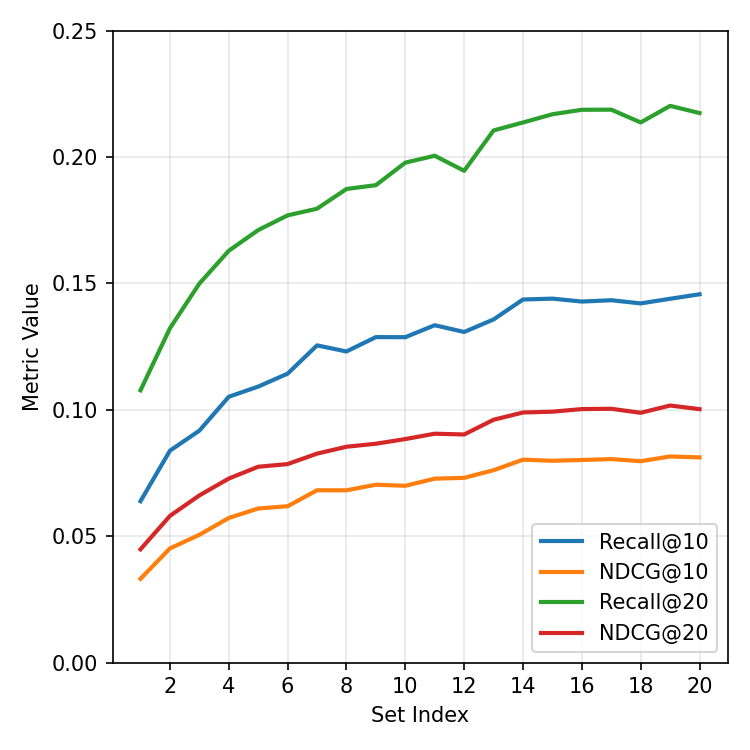}
        \caption{Games, Qwen3-Embedding-8B}
        \label{fig:app_item_sub3}
    \end{subfigure}

    \begin{subfigure}{0.32\textwidth}
        \includegraphics[width=\textwidth]{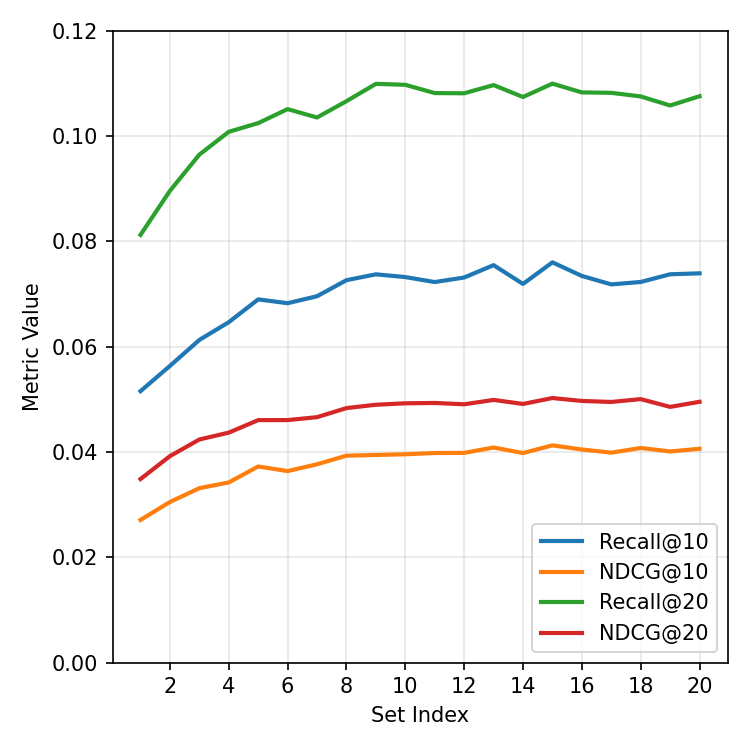}
        \caption{Toys, NV-Embed-v2}
        \label{fig:app_item_sub4}
    \end{subfigure}
    \hfill
    \begin{subfigure}{0.32\textwidth}
        \includegraphics[width=\textwidth]{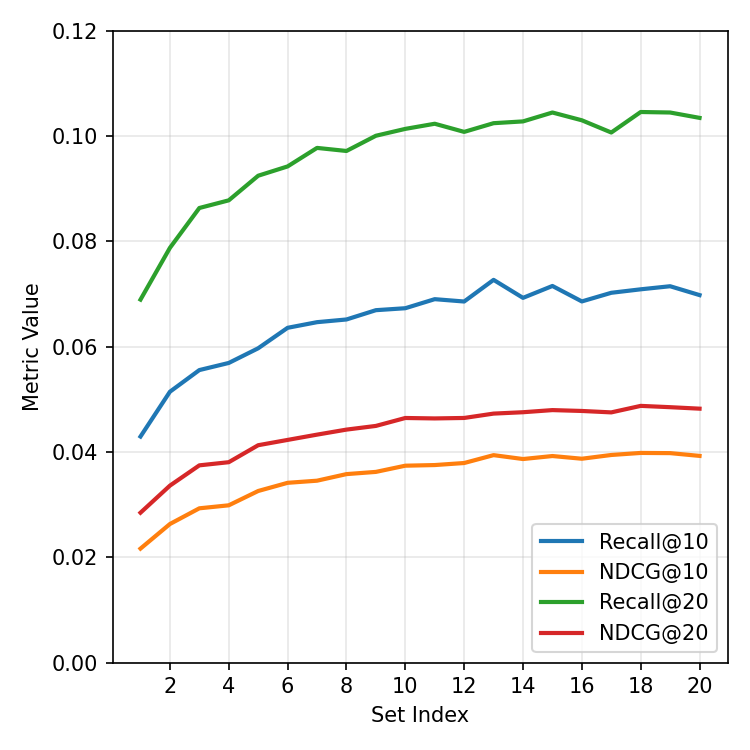}
        \caption{Toys, LLaMA-3.2-3B}
        \label{fig:app_item_sub5}
    \end{subfigure}
    \hfill
    \begin{subfigure}{0.32\textwidth}
        \includegraphics[width=\textwidth]{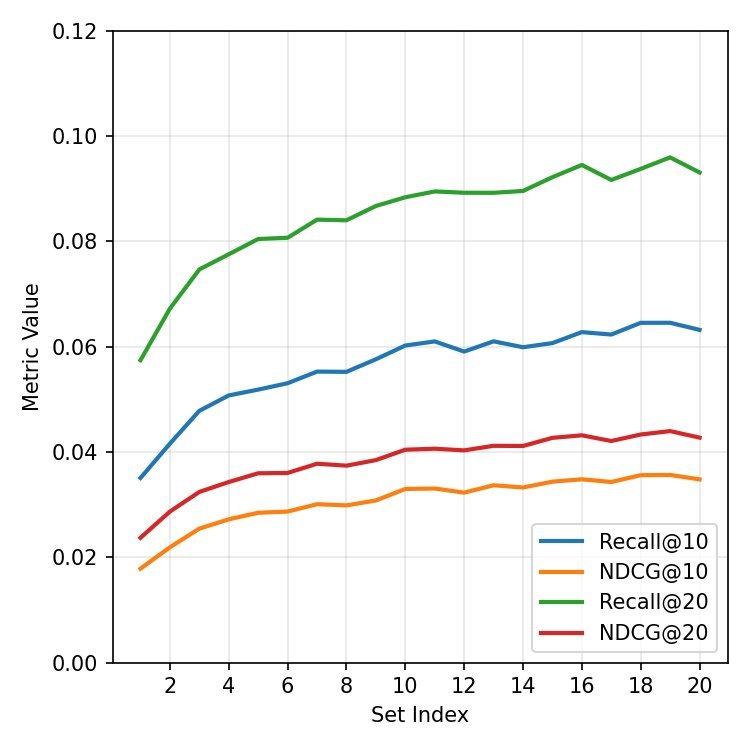}
        \caption{Toys, Qwen3-Embedding-8B}
        \label{fig:app_item_sub6}
    \end{subfigure}
    \caption{Item-only control experiment on Games and Toys across three LLM encoders. Performance consistently improves as item-semantic bands with progressively smaller singular values are included.}
    \Description{Six plots showing item-only semantic spectral utility on Games and Toys across NV-Embed-v2, LLaMA-3.2-3B, and Qwen3-Embedding-8B.}
    \label{fig:app_item_only}
\end{figure*}

To determine whether the semantic trend in Figure~\ref{fig:spectral_discrepancy} depends on constructing a user representation from the user's interaction history, we perform an item-only control experiment. We retain the item-side semantic matrix $\mathbf{S}_I$ produced by the frozen LLM encoder, but do not use $\mathbf{S}_U$. In particular, we compute an SVD of the item matrix alone:
\begin{equation}
\mathbf{S}_I = \mathbf{U}_I \operatorname{diag}(\boldsymbol{\sigma}_I) \mathbf{V}_I^\top,
\qquad \sigma_{I,1} \geq \cdots \geq \sigma_{I,r_I} > 0.
\end{equation}
We divide the ordered components into $B_{\mathrm{item}}=20$ contiguous groups of approximately equal size,
\begin{equation}
\mathbf{U}_I = [\mathbf{U}_I^{(1)}\|\cdots\|\mathbf{U}_I^{(B_{\mathrm{item}})}],\qquad
\mathbf{V}_I = [\mathbf{V}_I^{(1)}\|\cdots\|\mathbf{V}_I^{(B_{\mathrm{item}})}],
\end{equation}
and construct the cumulative item representation for set index $b$ as
\begin{equation}
\bar{\mathbf{S}}_I^{(1:b)} = \sum_{j=1}^{b} \mathbf{U}_I^{(j)}{\mathbf{V}_I^{(j)}}^\top,
\qquad b\in\{1,\ldots,B_{\mathrm{item}}\}.
\end{equation}
As in Section~\ref{sec:spectral_discrepancy}, the singular values are omitted in this reconstruction so that each retained singular component has equal weight. Thus, increasing $b$ adds progressively smaller, non-principal item-semantic components without changing their relative scale through $\boldsymbol{\sigma}_I$.

For every set index $b$, we train a fresh model with a trainable user-ID embedding $\mathbf{E}_U^{(b)}\in\mathbb{R}^{|\mathcal{U}|\times d_c}$ and a trainable semantic projection $\mathbf{W}^{(b)}\in\mathbb{R}^{d_s\times d_c}$. The item representation and predicted interaction matrix are
\begin{equation}
\mathbf{H}_I^{(b)} = \bar{\mathbf{S}}_I^{(1:b)}\mathbf{W}^{(b)},\qquad
\hat{\mathbf{R}}_{\mathrm{item}}^{(b)} = \mathbf{E}_U^{(b)}{\mathbf{H}_I^{(b)}}^\top,
\end{equation}
so that $\hat r_{ui}^{(b)}$ is the $(u,i)$ entry of $\hat{\mathbf{R}}_{\mathrm{item}}^{(b)}$. The model is optimized with the same BPR objective and data split as in Section~\ref{sec:experiments}:
\begin{equation}
\mathcal{L}_{\mathrm{BPR}}^{(b)} = -\sum_{(u,i,j)}
\log\!\left(\operatorname{sigmoid}\!\left(\hat r_{ui}^{(b)}-\hat r_{uj}^{(b)}\right)\right),
\end{equation}
where $i$ is an observed training item and $j$ is a sampled unobserved item. Consequently, the only semantic input is the item matrix $\mathbf{S}_I$; the user side is represented by collaborative trainable IDs rather than by an LLM-generated $\mathbf{S}_U$.

Figure~\ref{fig:app_item_only} shows that the resulting curves are consistent with the full user--item semantic analysis: for all three encoders, recommendation quality increases overall as additional item-semantic groups are included, and the full item spectrum performs better than the principal group alone. This control therefore supports the conclusion that the utility of non-principal semantic components is not solely caused by the user-profile construction from interaction histories.

\section{Overall Performance Comparison}
\label{app:overall_performance}
\noindent\textit{Complementing Section~\ref{sec:overall_performance}.}

Table~\ref{tab:app_overall_performance} reports the complete performance comparison across Games, Toys, and Books for all three LLM encoders (LLaMA-3.2-3B, NV-Embed-v2, and Qwen3-Embedding-8B), extending Table~\ref{tab:overall_performance} in Section~\ref{sec:overall_performance}.

\begin{table*}[htbp]
  \centering
  \caption{Overall performance comparison across three datasets with different semantic encoders. 
  ``$*$'' indicates statistical significance at $p<0.05$ for a two-tailed $t$-test against the best baseline.}
  \label{tab:app_overall_performance}
  \begin{subtable}[ht]{\textwidth}
    \centering
    \resizebox{\textwidth}{!}{%
      \begin{tabular}{l|cccc|cccc|cccc}
        \toprule
        \multirow{2}{*}{\textbf{Model}} 
        & \multicolumn{4}{c|}{\textbf{Games}} 
        & \multicolumn{4}{c|}{\textbf{Toys}} 
        & \multicolumn{4}{c}{\textbf{Books}} \\
        & R@10 & R@20 & N@10 & N@20 
        & R@10 & R@20 & N@10 & N@20 
        & R@10 & R@20 & N@10 & N@20 \\
        \midrule
        \multicolumn{13}{c}{Interaction-only recommendation models} \\
        \midrule
        LightGCN & 0.1404 & 0.2169 & 0.0770 & 0.0973 & 0.0505 & 0.0789 & 0.0282 & 0.0357 & 0.0911 & 0.1410 & 0.0523 & 0.0656 \\
        SimGCL & 0.1536 & 0.2223 & 0.0831 & 0.1016 & 0.0597 & 0.0915 & 0.0332 & 0.0417 & 0.1104 & 0.1611 & 0.0646 & 0.0788 \\
        SGFCF & 0.1873 & 0.2651 & 0.1085 & 0.1291 & 0.1005 & 0.1355 & 0.0607 & 0.0700 & 0.1743 & 0.2316 & 0.1099 & 0.1260 \\
        \midrule
        \multicolumn{13}{c}{LLM-enhanced models} \\
        \midrule
        L$^3$AE & 0.1878 & 0.2641 & 0.1083 & 0.1288 & 0.1139 & 0.1540 & 0.0674 & 0.0781 & 0.1797 & 0.2376 & 0.1137 & 0.1300 \\
        RLMRec-Con & 0.1667 & 0.2403 & 0.0911 & 0.1107 & 0.0686 & 0.1013 & 0.0388 & 0.0475 & 0.1165 & 0.1633 & 0.0692 & 0.0824 \\
        RLMRec-Gen & 0.1627 & 0.2379 & 0.0915 & 0.1114 & 0.0729 & 0.1093 & 0.0405 & 0.0502 & 0.1155 & 0.1631 & 0.0681 & 0.0814 \\
        AlphaRec & 0.1401 & 0.2108 & 0.0791 & 0.0952 & 0.0720 & 0.1050 & 0.0400 & 0.0488 & 0.1085 & 0.1578 & 0.0627 & 0.0765 \\
        ACE & 0.1589 & 0.2362 & 0.0872 & 0.1079 & 0.0627 & 0.0953 & 0.0334 & 0.0420 & 0.1067 & 0.1513 & 0.0624 & 0.0749 \\
        AlphaFuse & 0.1646 & 0.2423 & 0.0916 & 0.1126 & 0.0609 & 0.0929 & 0.0332 & 0.0417 & 0.1231 & 0.1747 & 0.0728 & 0.0872 \\
        Full Decoupling
        & 0.1706 & 0.2443 & 0.0948 & 0.1146
        & 0.1018 & 0.1416 & 0.0595 & 0.0701
        & 0.1388 & 0.1905 & 0.0825 & 0.0970 \\
        \hline
        UniSpecRec
        & \textbf{0.1922}$^{*}$
        & \textbf{0.2787}$^{*}$
        & \textbf{0.1137}$^{*}$
        & \textbf{0.1370}$^{*}$
        & \textbf{0.1155}$^{*}$
        & \textbf{0.1563}$^{*}$
        & \textbf{0.0694}$^{*}$
        & \textbf{0.0803}$^{*}$
        & \textbf{0.1816}$^{*}$
        & \textbf{0.2419}$^{*}$
        & \textbf{0.1143}
        & \textbf{0.1312}$^{*}$ \\
        \bottomrule
      \end{tabular}%
    }
    \caption{LLaMA-3.2-3B}
    \label{tab:sub_llama}
  \end{subtable}

  \begin{subtable}[ht]{\textwidth}
    \centering
    \resizebox{\textwidth}{!}{%
      \begin{tabular}{l|cccc|cccc|cccc}
        \toprule
        \multirow{2}{*}{\textbf{Model}} 
        & \multicolumn{4}{c|}{\textbf{Games}} 
        & \multicolumn{4}{c|}{\textbf{Toys}} 
        & \multicolumn{4}{c}{\textbf{Books}} \\
        & R@10 & R@20 & N@10 & N@20 
        & R@10 & R@20 & N@10 & N@20 
        & R@10 & R@20 & N@10 & N@20 \\
        \midrule
        \multicolumn{13}{c}{Interaction-only recommendation models} \\
        \midrule
        LightGCN & 0.1404 & 0.2169 & 0.0770 & 0.0973 & 0.0505 & 0.0789 & 0.0282 & 0.0357 & 0.0911 & 0.1410 & 0.0523 & 0.0656 \\
        SimGCL & 0.1536 & 0.2223 & 0.0831 & 0.1016 & 0.0597 & 0.0915 & 0.0332 & 0.0417 & 0.1104 & 0.1611 & 0.0646 & 0.0788 \\
        SGFCF & 0.1873 & 0.2651 & 0.1085 & 0.1291 & 0.1005 & 0.1355 & 0.0607 & 0.0700 & 0.1743 & 0.2316 & 0.1099 & 0.1260 \\
        \midrule
        \multicolumn{13}{c}{LLM-enhanced models} \\
        \midrule
        L$^3$AE & \textbf{0.1966} & 0.2737 & 0.1128 & 0.1335 & \textbf{0.1168} & 0.1573 & 0.0701 & 0.0810 & \textbf{0.1818} & 0.2409 & \textbf{0.1151} & 0.1315 \\
        RLMRec-Con & 0.1626 & 0.2385 & 0.0902 & 0.1104 & 0.0691 & 0.1039 & 0.0381 & 0.0473 & 0.1153 & 0.1605 & 0.0692 & 0.0820 \\
        RLMRec-Gen & 0.1670 & 0.2391 & 0.0944 & 0.1136 & 0.0826 & 0.1212 & 0.0457 & 0.0559 & 0.1237 & 0.1742 & 0.0731 & 0.0873 \\
        AlphaRec & 0.1740 & 0.2536 & 0.0991 & 0.1204 & 0.0804 & 0.1206 & 0.0442 & 0.0547 & 0.1223 & 0.1698 & 0.0722 & 0.0856 \\
        ACE & 0.1715 & 0.2471 & 0.0972 & 0.1175 & 0.0711 & 0.1072 & 0.0389 & 0.0484 & 0.1129 & 0.1572 & 0.0671 & 0.0797 \\
        AlphaFuse & 0.1708 & 0.2552 & 0.0966 & 0.1191 & 0.0726 & 0.1083 & 0.0394 & 0.0488 & 0.1164 & 0.1655 & 0.0691 & 0.0828 \\
        Full Decoupling
        & 0.1752 & 0.2532 & 0.0998 & 0.1207
        & 0.1068 & 0.1475 & 0.0618 & 0.0726
        & 0.1365 & 0.1908 & 0.0815 & 0.0967 \\
        \hline
        UniSpecRec
        & 0.1934
        & \textbf{0.2823}$^{*}$
        & \textbf{0.1143}$^{*}$
        & \textbf{0.1381}$^{*}$
        & 0.1160
        & \textbf{0.1601}$^{*}$
        & \textbf{0.0704}
        & \textbf{0.0821}$^{*}$
        & 0.1812
        & \textbf{0.2426}$^{*}$
        & 0.1144
        & \textbf{0.1317} \\
        \bottomrule
      \end{tabular}%
    }
    \caption{NV-Embed-v2}
    \label{tab:sub_nvidia}
  \end{subtable}

  \begin{subtable}[ht]{\textwidth}
    \centering
    \resizebox{\textwidth}{!}{%
      \begin{tabular}{l|cccc|cccc|cccc}
        \toprule
        \multirow{2}{*}{\textbf{Model}} 
        & \multicolumn{4}{c|}{\textbf{Games}} 
        & \multicolumn{4}{c|}{\textbf{Toys}} 
        & \multicolumn{4}{c}{\textbf{Books}} \\
        & R@10 & R@20 & N@10 & N@20 
        & R@10 & R@20 & N@10 & N@20 
        & R@10 & R@20 & N@10 & N@20 \\
        \midrule
        \multicolumn{13}{c}{Interaction-only recommendation models} \\
        \midrule
        LightGCN & 0.1404 & 0.2169 & 0.0770 & 0.0973 & 0.0505 & 0.0789 & 0.0282 & 0.0357 & 0.0911 & 0.1410 & 0.0523 & 0.0656 \\
        SimGCL & 0.1536 & 0.2223 & 0.0831 & 0.1016 & 0.0597 & 0.0915 & 0.0332 & 0.0417 & 0.1104 & 0.1611 & 0.0646 & 0.0788 \\
        SGFCF & 0.1873 & 0.2651 & 0.1085 & 0.1291 & 0.1005 & 0.1355 & 0.0607 & 0.0700 & 0.1743 & 0.2316 & 0.1099 & 0.1260 \\
        \midrule
        \multicolumn{13}{c}{LLM-enhanced models} \\
        \midrule
        L$^3$AE & 0.1822 & 0.2625 & 0.1042 & 0.1257 & 0.1060 & 0.1441 & 0.0627 & 0.0729 & 0.1790 & 0.2356 & \textbf{0.1139} & 0.1299 \\
        RLMRec-Con & 0.1646 & 0.2371 & 0.0908 & 0.1103 & 0.0674 & 0.0995 & 0.0382 & 0.0467 & 0.1154 & 0.1631 & 0.0691 & 0.0824 \\
        RLMRec-Gen & 0.1486 & 0.2251 & 0.0833 & 0.1037 & 0.0746 & 0.1100 & 0.0414 & 0.0508 & 0.1206 & 0.1692 & 0.0713 & 0.0850 \\
        AlphaRec & 0.1671 & 0.2443 & 0.0926 & 0.1140 & 0.0687 & 0.1018 & 0.0375 & 0.0463 & 0.1228 & 0.1713 & 0.0732 & 0.0869 \\
        ACE & 0.1655 & 0.2445 & 0.0916 & 0.1126 & 0.0677 & 0.1029 & 0.0371 & 0.0464 & 0.1132 & 0.1601 & 0.0664 & 0.0797 \\
        AlphaFuse & 0.1639 & 0.2430 & 0.0915 & 0.1126 & 0.0686 & 0.1016 & 0.0380 & 0.0468 & 0.1160 & 0.1602 & 0.0688 & 0.0813 \\
        Full Decoupling
        & 0.1611 & 0.2472 & 0.0912 & 0.1142
        & 0.0873 & 0.1266 & 0.0510 & 0.0601
        & 0.1342 & 0.1869 & 0.0802 & 0.0947 \\
        \hline
        UniSpecRec
        & \textbf{0.1905}$^{*}$
        & \textbf{0.2741}$^{*}$
        & \textbf{0.1105}$^{*}$
        & \textbf{0.1327}$^{*}$
        & \textbf{0.1098}$^{*}$
        & \textbf{0.1496}$^{*}$
        & \textbf{0.0656}$^{*}$
        & \textbf{0.0761}$^{*}$
        & \textbf{0.1810}$^{*}$
        & \textbf{0.2420}$^{*}$
        & \textbf{0.1139}
        & \textbf{0.1310}$^{*}$ \\
        \bottomrule
      \end{tabular}%
    }
    \caption{Qwen3-Embedding-8B}
    \label{tab:sub_qwen}
  \end{subtable}
\end{table*}

\begin{figure}[t]
    \centering
    \begin{subfigure}{0.45\textwidth}
        \centering
        \includegraphics[width=\linewidth]{imgs/spectral_filters/llama.png}
        \subcaption{LLaMA-3.2-3B}
        \label{subfig:llama_filter}
    \end{subfigure}

    \begin{subfigure}{0.45\textwidth}
        \centering
        \includegraphics[width=\linewidth]{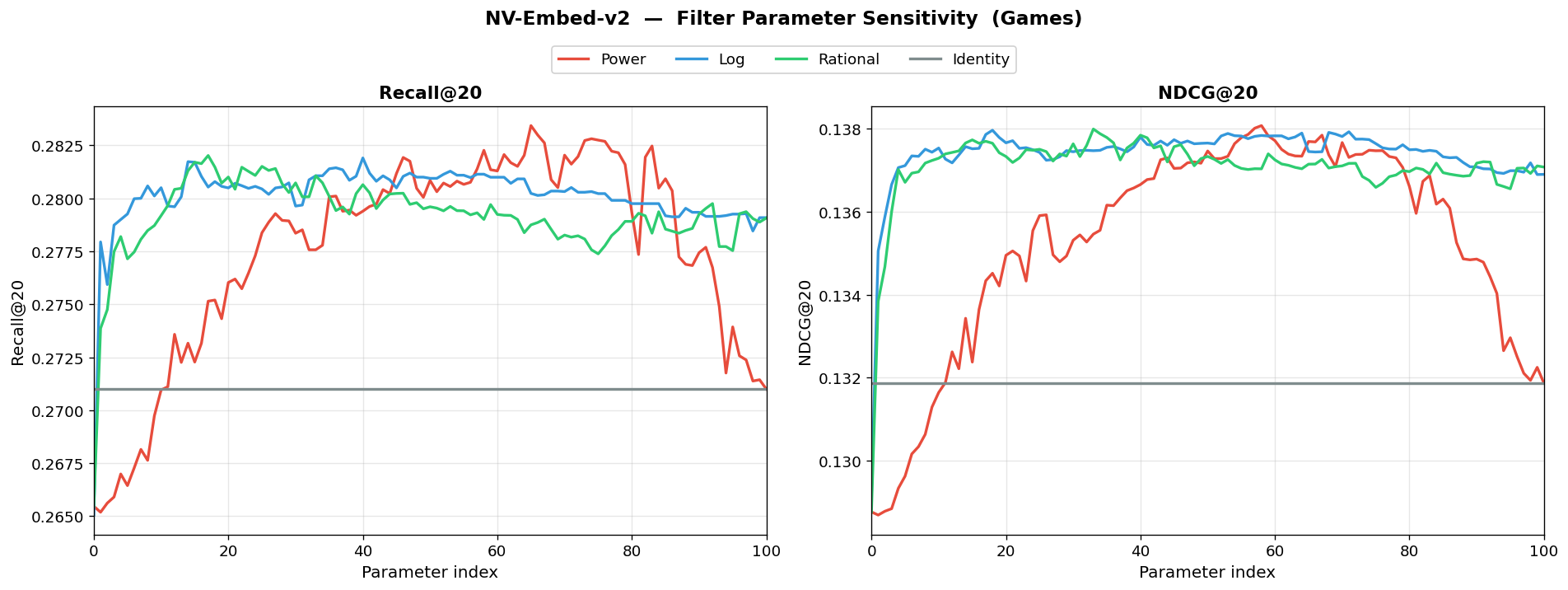}
        \subcaption{NV-Embed-v2}
        \label{subfig:nvidia_filter}
    \end{subfigure}

    \begin{subfigure}{0.45\textwidth}
        \centering
        \includegraphics[width=\linewidth]{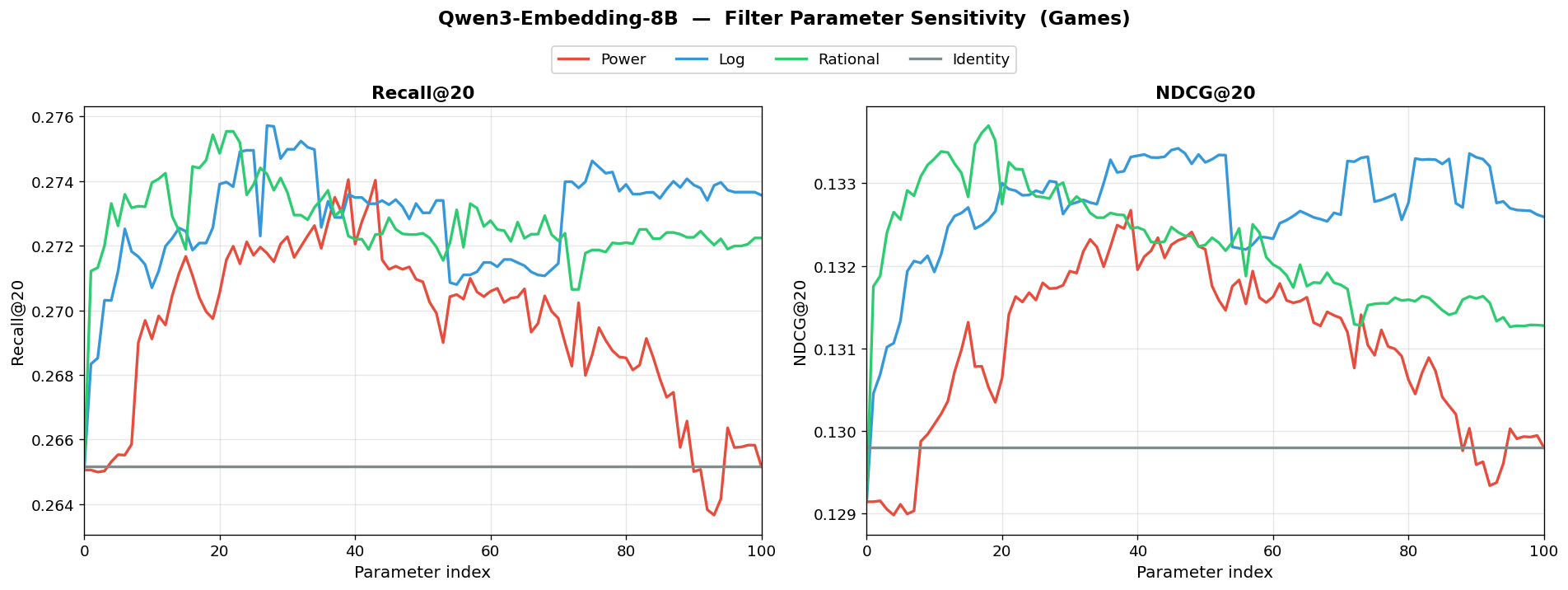}
        \subcaption{Qwen3-Embedding-8B}
        \label{subfig:qwen_filter}
    \end{subfigure}
    \caption{Performance comparison of different spectral filter designs across three LLM encoders on Games.}
    \label{fig:app_spectral_filters}
\end{figure}

\section{Model Parameter Analysis}
\label{app:param_analysis}
\noindent\textit{Complementing Section~\ref{sec:model_analysis}.}

\subsection{Spectral Filter Design}

The semantic branch is obtained from the SVD of the joint semantic matrix
$\mathbf{S}=\mathbf{U}_S\operatorname{diag}(\boldsymbol{\sigma}_s)\mathbf{V}_S^\top$.
For a filter $f(\cdot)$, its prediction can be written in the factorized form
\begin{equation}
\hat{\mathbf{R}}_S
=\mathbf{U}_{S_U}\operatorname{diag}\!\left(f^2(\boldsymbol{\sigma}_s)\right)
\mathbf{U}_{S_I}^{\top},
\end{equation}
where $\mathbf{U}_{S_U}$ and $\mathbf{U}_{S_I}$ are the user and item row
partitions of $\mathbf{U}_S$. We compare the same spectral filters as in Section~\ref{sec:model_analysis}: the identity filter $f(\sigma)=\sigma$, the power filter $f(\sigma)=\sigma^p$ ($p\in[0,1]$), the log filter $f_{\mathrm{log}}(\sigma)=\log(1+\gamma\sigma/\sigma_{\max})$ ($\gamma\in[0,250]$, $\sigma_{\max}=\max_k\sigma_k$), the rational filter $f_{\mathrm{rat}}(\sigma)=\frac{\sigma/\sigma_{\max}}{\sigma/\sigma_{\max}+\beta}$ ($\beta\in[0,0.25]$), and a learnable filter $f_{\mathrm{train}}(\sigma)=\operatorname{MLP}(\sigma)$ optimized with the BPR objective on the validation split.

Figure~\ref{fig:app_spectral_filters} summarizes the sensitivity curves on Games for
all three encoders. The hand-designed filters are generally stronger than the
learnable filter, and the identity baseline is consistently among the weakest
settings. The power filter is comparatively stable over its search range,
whereas the log and rational filters exhibit larger fluctuations. This behavior
is consistent with the sharply concentrated semantic spectrum: a moderate power
can compensate for the small magnitude of non-principal components without
allowing the filter to dominate the collaborative signal. The result motivates
using the power filter as the default parameterization in UniSpecRec.

\subsection{Balancing Coefficient and Filter Exponent}
After filtering, UniSpecRec combines the two prediction branches as
\begin{equation}
\hat{\mathbf{R}}
=(1-\alpha)\hat{\mathbf{R}}_{\mathrm{CF}}
+\alpha\hat{\mathbf{R}}_S,
\qquad \alpha\in[0,1].
\end{equation}
Here, $p$ controls the relative weighting of semantic spectral components,
whereas $\alpha$ controls the contribution of the entire semantic prediction
relative to the collaborative prediction. Thus, the two parameters address
different sources of variation and should not be interpreted as interchangeable:
changing $p$ reshapes $\hat{\mathbf{R}}_S$, while changing $\alpha$ changes the
fusion point between $\hat{\mathbf{R}}_{\mathrm{CF}}$ and $\hat{\mathbf{R}}_S$.

\begin{figure}[t]
    \centering
    \begin{subfigure}{0.4\textwidth}
        \centering
        \includegraphics[width=\linewidth]{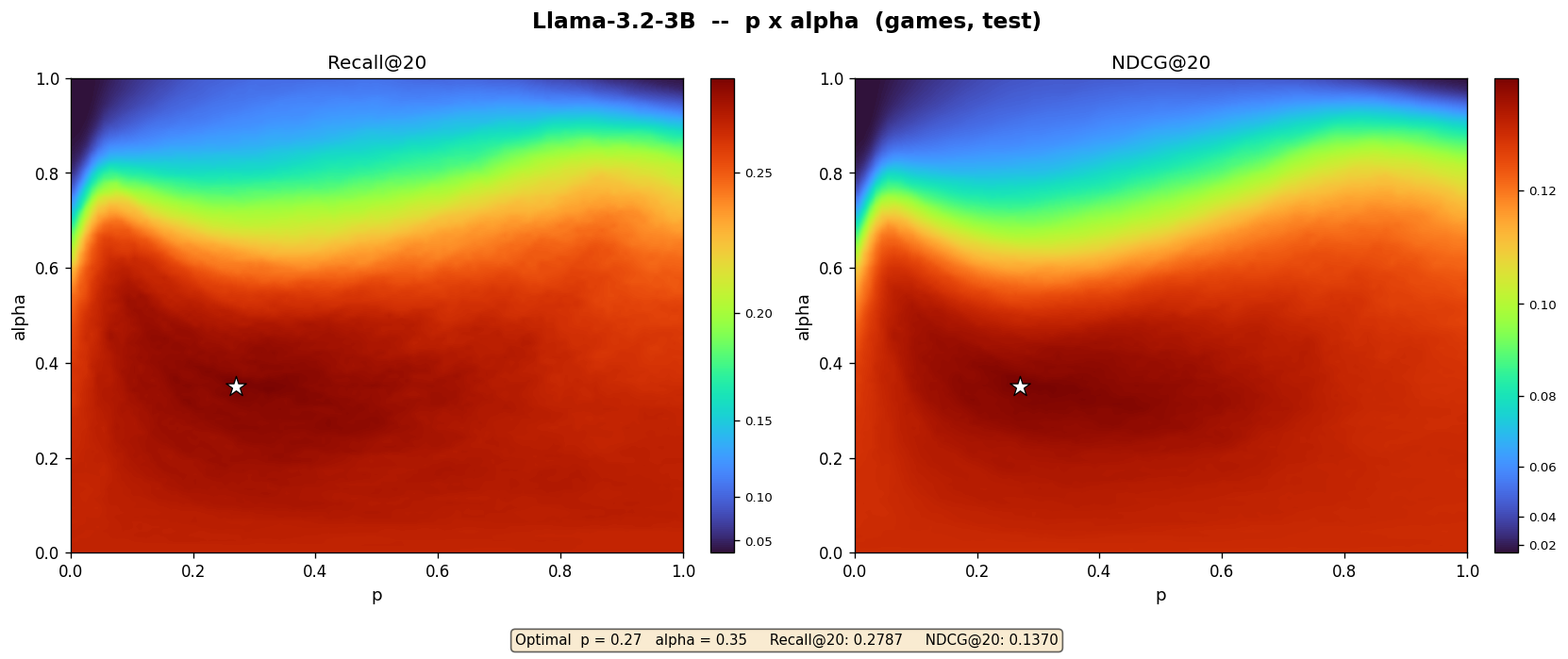}
        \subcaption{LLaMA-3.2-3B}
        \label{subfig:llama_heatmap}
    \end{subfigure}

    \begin{subfigure}{0.4\textwidth}
        \centering
        \includegraphics[width=\linewidth]{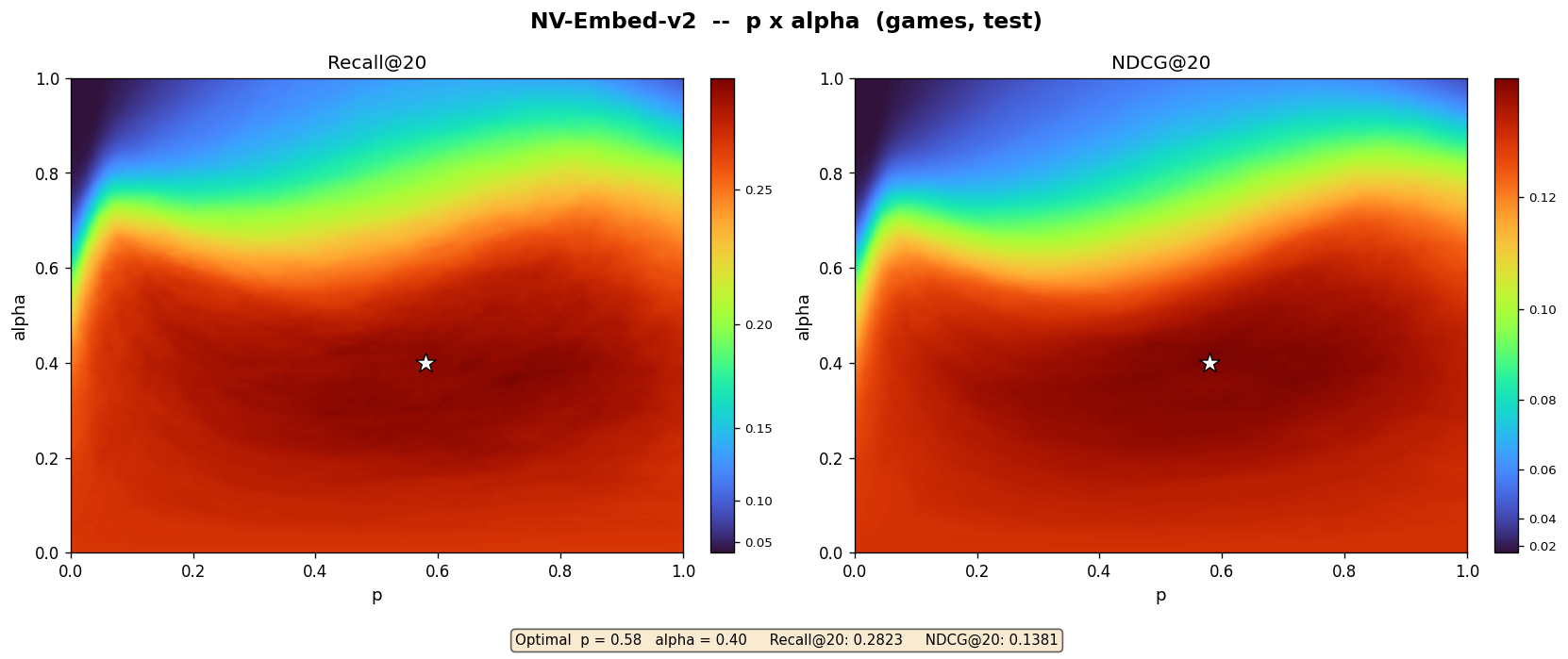}
        \subcaption{NV-Embed-v2}
        \label{subfig:nvidia_heatmap}
    \end{subfigure}

    \begin{subfigure}{0.4\textwidth}
        \centering
        \includegraphics[width=\linewidth]{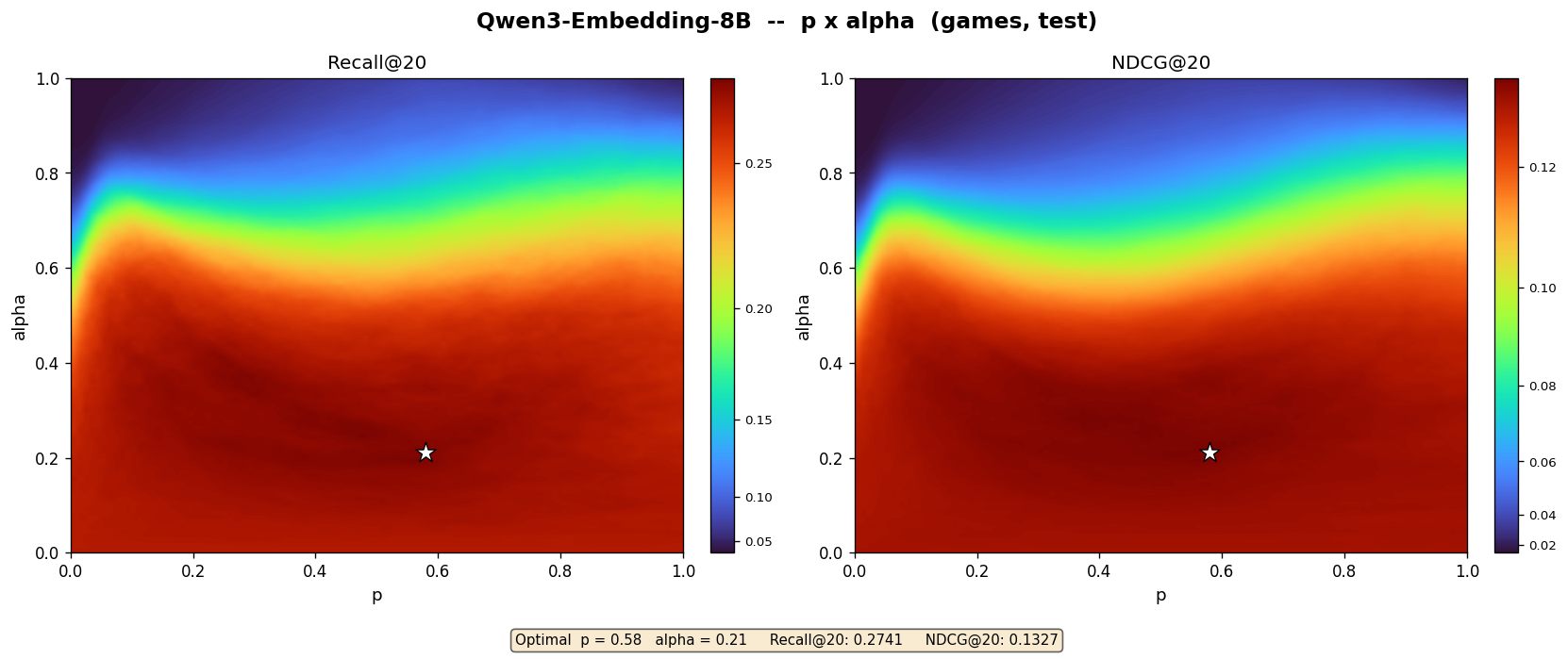}
        \subcaption{Qwen3-Embedding-8B}
        \label{subfig:qwen_heatmap}
    \end{subfigure}

    \caption{Performance heatmap of balancing coefficient $\alpha$ and filter parameter $p$ across three LLM encoders on Games.}
    \label{fig:p_alpha_heatmap_full}
\end{figure}

Figure~\ref{fig:p_alpha_heatmap_full} shows their joint sensitivity across the three encoders on Games. The best configurations use intermediate values of $p$, rather than either endpoint, indicating that retaining the semantic spectrum without reweighting is suboptimal and that aggressively flattening it can also be harmful. The preferred values of $\alpha$ are likewise interior: the semantic branch improves the collaborative predictor when it is given a substantial but not dominant weight. Although the exact optimum varies with the encoder, the high-performing region is broad, which supports selecting $p$ and $\alpha$ with a modest validation sweep instead of relying on a finely tuned, encoder-specific setting. These observations are consistent with the dataset-level validation selections reported in Table~\ref{tab:param_analysis}.

\FloatBarrier

\end{document}